%% file: src_low_res.tex
\documentclass[10pt,aps,prc,floatfix,twocolumn,nofootinbib]{revtex4-1}

\usepackage{amsfonts}
\usepackage{amsmath}
\usepackage{amssymb}
\usepackage{bbold}
\usepackage{bm}
\usepackage{cellspace}
\usepackage{color}
\usepackage{enumerate}
\usepackage{graphicx}
\usepackage{physics} 
\usepackage[figuresright]{rotating}
\usepackage{siunitx}
\usepackage[caption=false]{subfig} 
\usepackage{simpler-wick}
\usepackage{xfrac}
\usepackage{hyperref} 

\graphicspath{{figures_low_res/},{figures/}} 

\input{macros}

\begin{document}

\title{Short-range correlation physics at low RG resolution}

\author{A.~J.~Tropiano$^{1}$, S.~K.~Bogner$^{2}$, R.~J.~Furnstahl$^{1}$}

\affiliation{%
$^1$\mbox{Department of Physics, The Ohio State University, Columbus, OH 43210, USA}  \\
$^2$\mbox{Facility for Rare Isotope Beams and Department of Physics and Astronomy,}  \\
\mbox{Michigan State University, East Lansing, MI 48824, USA}
}

\date{\today}

\begin{abstract}
Recent experiments have succeeded in isolating processes for which short-range correlation (SRC) physics is dominant and well accounted for by SRC phenomenology.
But an alternative and compelling picture emerges from renormalization group (RG) evolution to low RG resolution.
At high RG resolution, SRCs are identified as components in the nuclear wave function with relative pair momenta greater than the Fermi momentum.
Scale separation results in wave-function factorization that can be exploited with phenomenologies such as the generalized contact formalism or the low-order correlation operator approximation. 
Evolution to lower resolution shifts SRC physics from nuclear structure to the reaction operators without changing the measured observables.
We show how the features of SRC phenomenology manifested at high RG resolution are cleanly identified 
at low RG resolution
using simple two-body operators and local-density approximations with uncorrelated wave functions,
all of which can be systematically generalized.
We verify that the experimental consequences to date follow directly at low resolution from well-established properties of nucleon-nucleon interactions such as the tensor force.
Thus the RG reconciles the contrasting pictures of the same experiment and shows how to get correct results using wave functions without SRC components.
Our demonstration has implications for the analysis of knock-out reactions for which SRC physics is not cleanly isolated.
\end{abstract}

\maketitle

\newpage

\section{Introduction} \label{sec:introduction}

Short-range correlations (SRCs) in atomic nuclei are usually identified as components of the nuclear wave function with nucleon pair momenta well above the Fermi momentum~\cite{Hen:2016kwk}. 
There has long been an apparent need for such SRCs to account for measured cross sections~\cite{Brueckner:1955zzd}, but direct evidence has been nebulous until recent experiments succeeded in cleanly isolating this physics~\cite{Hen:2016kwk,Korover:2014dma,Hen:2014nza,Duer:2018sby,Duer:2018sxh,Schmookler:2019nvf,Cruz-Torres:2020uke,Schmidt:2020kcl,CLAS:2020rue}. 
SRC phenomenologies have been developed that account for the observations, but they seem to be at odds with successful descriptions of nuclear structure such as the shell model that do not feature explicit short-range structure in the nuclear wave function. 
The application of the renormalization group (RG) can make sense of this conflict. 
RG methods are used to analyze critical phenomena in condensed matter~\cite{Amit:2005} and evolve the strong coupling and parton distributions in high-energy quantum chromodynamics (QCD)~\cite{RevModPhys.67.157,peskin1995introduction}.
Applied to nuclei, the RG shows how SRC physics is manifested differently at varying resolution scales. 
Here we will illustrate how the RG can bridge low- and high-resolution treatments of the same experiment, shedding light on the implications of SRC physics and on long-standing discrepancies between theory and experiment~\cite{Aumann:2020tcq}.

To avoid misunderstanding, we must from the beginning distinguish between experimental resolution and RG resolution as manifested in nuclear applications.
The experimental resolution is set by the momentum of the probe, with the resolving power limited to distances of order the corresponding wavelength.
The RG resolution is also determined by a limiting wavelength, which is set not by external kinematics but by the choice of decoupling scale in the RG-evolved Hamiltonian.
This scale dictates the minimum wavelength or maximum momentum available for the wave functions of low-energy states (and the nuclear ground states in particular).
A high-RG-resolution description has a ``hard'' Hamiltonian that mixes high-momentum components into low-energy states, i.e., SRCs. 
A low-RG-resolution description has a ``soft'' Hamiltonian for which the ground-state wave function is closer to the mean-field limit with the largest momenta not far from the Fermi momentum $\kF$.
We emphasize that either of these descriptions (or a continuum of intermediate RG resolutions) can be applied to the same experiment.

\begin{figure}[tbh]
    \centering
    (a)\includegraphics[width=0.43\columnwidth]{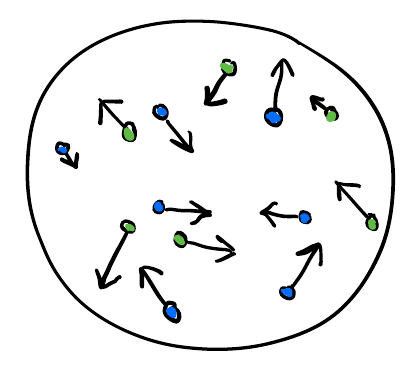}~~%
    (b)\includegraphics[width=0.44\columnwidth]{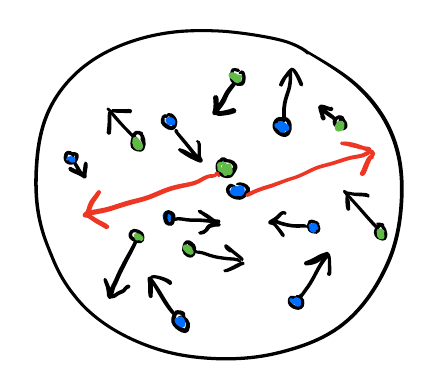}
    \caption{Cartoon snapshots of a nucleus at (a) low-RG  and (b) high-RG resolutions. 
    The back-to-back nucleons at high-RG resolution are an SRC pair with small center-of-mass momentum.
    }
    \label{fig:cartoon_nuclei}
\end{figure}

\begin{figure}[tbh]
    \centering
    \includegraphics[width=0.9\columnwidth]{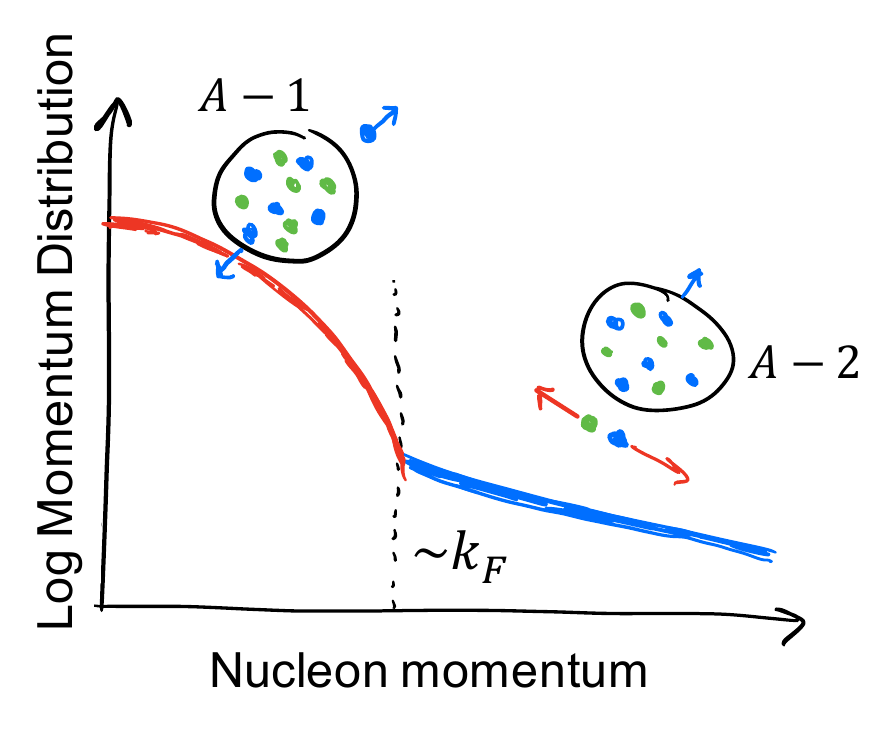}
    \caption{Cartoon of a nuclear single-particle momentum distribution at high RG resolution.
    The region below $\kF$ is a Fermi sea of nucleons.
    Above $\kF$ the distribution is dominated by SRC pairs, with center-of-mass momenta of order $\kF$.
    At low RG resolution, the upper branch is greatly suppressed.
    }
    \label{fig:cartoon_mom_dist}
\end{figure}

For visualization, cartoon pictures of these resolutions are shown in Figs.~\ref{fig:cartoon_nuclei} and \ref{fig:cartoon_mom_dist}.
Although only qualitative, such pictures give intuition for interpreting experiments.
In Fig.~\ref{fig:cartoon_nuclei} we visualize configurations that might be sampled at different RG resolutions. Only at high-RG resolution do we find a high-momentum pair.
In practice we associate the high-resolution picture with local phenomenological Hamiltonians, in particular the Argonne V18 (AV18) potential~\cite{Wiringa:1994wb} and its associated three-body forces~\cite{Carlson:1983kq,Pieper:2001ap}, which have been featured in most of the recent analyses of SRC experiments~\cite{Hen:2016kwk}. 
AV18 describes nucleon-nucleon observables within experimental accuracy up to the inelastic threshold and the elastic part of the cross section reasonably well up to much higher energies.
The AV18 momentum distributions in nuclei extend well above 4\,fm$^{-1}$ or 800\,MeV (in units where $\hbar = c = 1$), as caricatured in Fig.~\ref{fig:cartoon_mom_dist} (see also the figures in Sec.~\ref{sec:low_resolution}).
For concreteness we associate the low-resolution picture with the shell model (configuration interaction), either in its phenomenological form~\cite{Heyde:2004zz,Ring:2005} or derived from ab initio methods such as coupled cluster or the in-medium similarity RG~\cite{Hergert:2020bxy}.
In this picture, nucleon momentum distributions die off rapidly above $\kF$.

The smoking gun experiments for SRCs in the high-RG-resolution picture come in two varieties: inclusive reactions with carefully chosen kinematics~\cite{Egiyan:2003vg,Egiyan:2005hs,Fomin:2011ng,Ye:2017mvo,Nguyen:2020mgo} and experiments featuring the knock-out and detection of two high-momentum nucleons~\cite{Hen:2014nza,Higinbotham:2014xna,Korover:2014dma,Duer:2018sxh,Duer:2018sby,Schmookler:2019nvf,Schmidt:2020kcl}.
These experiments manifest key features of high-resolution SRC phenomenology~\cite{Hen:2016kwk}:
\begin{enumerate}
  \item \textbf{Universal high-momentum nucleon distributions.}
  Ratios of inclusive cross sections at selective kinematics (high $Q^2$ and $x \gtrsim 1.4$--1.5) for different nuclei show plateaus, implying that momentum distributions at high momenta have the same shape for all nuclei.
  The relative height of the plateau defines the SRC scaling factor $a_2$ and is related to the relative probability of finding a two-nucleon SRC in the nuclei (e.g., see Ref.~\cite{Ryckebusch:2019oya}).

   \item \textbf{Kinematics of the knocked-out nucleons.}
   SRC pairs are in S-waves with relative momentum larger than $\kF$ and total momentum of order or less than $\kF$.
   So when one of the paired nucleons has a direct one-body interaction (e.g., it absorbs a virtual photon), the other flies out almost back-to-back with respect to the original momentum in the lab frame.
   The experimental analysis implies that roughly 10--20\% of nucleons in the nucleus are members of an SRC pair.
   
  \item \textbf{Ratio of $np$ to $pp$ knocked-out pairs for intermediate relative momentum (300--500 MeV).}
  One might expect that the ratio of neutron-proton ($np$) to proton-proton ($pp$) pairs knocked out would be given by basic counting~\cite{Colle:2015ena}.
  In fact the ratio for intermediate momenta shows a strong dominance of $np$ pairs over $pp$ or $nn$ pairs.
  This is interpreted as resulting from the dominance of the nucleon-nucleon (NN) tensor interaction in this momentum regime, which generates \tripletS\ ($np$ only) but not \singletS\ SRC pairs.
  
   \item \textbf{Ratio of knock-out cross sections from neutron-rich nuclei compared to $N=Z$ nuclei.}
   The ratios imply that the high-momentum fraction of protons is greater for $N>Z$ than for $N=Z$.
   The SRC interpretation is that excess neutrons correlate with core protons, so that protons ``speed-up'' in neutron-rich nuclei.
   In contrast, the high-momentum fraction of neutrons remains constant.
   
  \item \textbf{Transition from $np$ dominance of SRC pairs to ratios expected from scalar counting.}
  As the momentum of the knocked-out pairs increases, the dominance of the tensor part of the NN interaction gives way to probing the repulsive core, which affects all pairs equally.
  Therefore the ratio of $np$ to $pp$ pairs relaxes toward the scalar limit.
  
\end{enumerate}
These effects are all consistent with two phenomenological approaches. 
The Generalized Contact Formalism (GCF) model uses a factorization ansatz for the nuclear many-body wavefunction applied to these high-momentum transfer processes~\cite{Weiss:2015mba,Alvioli:2016wwp,Weiss:2016obx,Weiss:2018tbu,Weiss:2021zyb}. 
The low-order correlation operator approximation (LCA)~\cite{Vanhalst:2014cqa,Ryckebusch:2018rct,Ryckebusch:2019oya} is used to compute how SRCs affect nuclear momentum distributions; it does so by shifting the SRC physics from the wave function to a correlation operator, which is in turn parametrized
(the LCA will be discussed further in Sec.~\ref{sec:low_resolution}).

The GCF embodies the pictures in Fig.~\ref{fig:cartoon_nuclei}(b) and Fig.~\ref{fig:cartoon_mom_dist}, with the nuclear wavefunction relevant for SRC physics events represented as a product of a short-distance piece that multiplies a mean-field piece.
When applied to the distribution of pair momenta in coordinate and momentum space (meaning the probability to find two nucleons within a distance $r$ or having a relative momentum $q$), these quantities take the factorized form:
\begin{align}
     \rho^A_{{\rm\scriptscriptstyle GCF},\alpha}(r)  &= C_{A}^{NN,\alpha} |\phi^\alpha_{NN}(r)|^2 \;,
     \label{eq:GCF_coordinate}
    \\
    n^A_{{\rm\scriptscriptstyle GCF},\alpha}(q) &= C_{A}^{NN,\alpha} |\widetilde \phi^\alpha_{NN}(q)|^2
    \;. \label{eq:GCF_momentum}
\end{align}
Here $A$ is the nucleon number (protons plus neutrons) and $\alpha$ is a spin index.
The common coefficient is called a ``contact'' in analogy to the quantity defined in cold atom physics~\cite{Tan2008contact1,Tan2008contact2}.
The two-body functions are extracted from the short-distance (high-momentum) dependence of the zero-energy solutions to the Schr\"odinger equation for a given NN interaction.%
\footnote{It would be misleading to call them wave functions, as pointed out originally in Ref.~\cite{Brueckner:1955zzd}.}

There is by now a substantial literature on applying the GCF phenomenology to SRC experiments~\cite{Weiss:2015mba,Weiss:2016obx,Weiss:2018tbu,Cruz-Torres:2019fum,Weiss:2021zyb,Pybus:2020itv,Schmidt:2020kcl}; we summarize only the major features here.
The $A$-independence of the $\phi$ functions immediately implies that both the short-distance and high-momentum dependence of the pair distributions in different nuclei will be universal.
This suggests that taking ratios will divide out model dependence, giving support to the early focus on cross section ratios of various sorts in experimental analyses (e.g., two-nucleon knock-out to integrated single-nucleon knock-out).
Ratios of contacts are scale and scheme independent, as validated by microscopic ab initio calculations using both hard and soft (although not fully soft) interactions.
Three-body effects are implicitly treated as negligible.

The success of the SRC pair interpretations and the GCF might seem to be conclusive evidence that the high-RG-resolution picture is the only correct one.
But in fact these observations can be explained as well by a very different picture, in which the Hamiltonian is at low-RG resolution and therefore soft, while the interaction of a virtual photon probe is not with a single nucleon that is part of an SRC pair but with two nucleons in the Fermi sea (i.e., a two-body current).
We claim that all of the same observables are reproduced in this alternative picture, with simpler calculations.
This might seem to be a nuclear Rashomon effect~\cite{Rashomon:2016}, in which different observers give contradictory interpretations of the same event.
The RG shows how these pictures can both describe the phenomena and how to continuously transform from one to the other.
We note that these contrasting pictures have been discussed in the nuclear RG literature for at least a decade~\cite{Anderson:2010aq,Bogner:2012zm,Furnstahl:2013dsa,Tropiano:2020zwb}, but this work has not made a substantial impact or when cited has often been misunderstood.
We hope the presentation here will prove to be more accessible.

In the remainder of this paper we flesh out the story, striving to provide intuitive explanations of how it plays out.
In Sec.~\ref{sec:historical} we give a selected history of the two alternative pictures of nuclei and the experiments that seemed to require SRCs in nuclei.
This takes us up to the present where the conflict persists; we reconcile the pictures in Sec.~\ref{sec:reconciliation} by means of the RG.
We provide low-resolution explanations of the SRC experiments, some directly and some by showing how the GCF phenomenology emerges from an RG treatment and the operator product expansion.
A basic second-quantized treatment that enables a systematic many-body treatment of the unitary RG approach, along with representative calculations supporting the Sec.~\ref{sec:reconciliation} explanations, is presented in Sec.~\ref{sec:low_resolution}.
Section~\ref{sec:implications} provides a selection of takeaways in the form of questions and answers, and some implications for other experiments where SRC physics may play a role.
Section~\ref{sec:summary} is a brief summary and guide to future work.

\section{Historical Antecedents} \label{sec:historical}

Since the 1950's, two apparently contradictory pictures of nuclear structure have been developed and successfully applied~\cite{preston1975structure}.
The first stems from the shell model of Goeppert Mayer and Jensen, originally a description of independent particles moving in a mean field.
Refinements from that era include the collective model of Bohr and Mottelson, which showed that the
vibrational and rotational excitations of nuclei can be described in terms of the time evolution of a self-consistent mean field.
This provided a unified description of single-particle and collective degrees of freedom in nuclei.
The key characteristic for our discussion is that the momentum distribution did not include high-momentum nucleons (that is, with momenta well above $\kF$), as depicted in Fig.~\ref{fig:cartoon_nuclei}(a).

In Ref.~\cite{Brueckner:1955zzd}, Brueckner et al.\ considered the possibility of nonlinear phenomena altering the strong short-range interaction evident in free space: ``\ldots the success of the shell model has often been assumed to indicate that the two-body forces in nuclear matter are in fact much weaker and long-ranged and can lead in an excellent approximation to a uniform Hartree field acting on the nucleons.''
They go on to soundly reject this conjecture by considering five high-energy reactions and concluding that the measured cross sections can only result from a nucleon momentum distribution with a significant tail: ``This momentum distribution differs substantially from that for the shell model of the nucleus and thus provides strong evidence for correlation in the nuclear ground-state wave function.''
This is the alternative picture in Figs.~\ref{fig:cartoon_nuclei}(b) and \ref{fig:cartoon_mom_dist}, which features in particular a component of the nuclear wave function consisting of \emph{pairs} of nucleons with large relative momentum but a center-of-mass momentum of order the Fermi momentum.
Brueckner et al.\ explained this picture as arising from the strong short-range repulsion in the nucleon-nucleon (NN) interaction, which was the accepted explanation for the NN S-wave phase shifts turning negative at high energies.
Thus the SRC was born.

A key takeaway from Ref.~\cite{Brueckner:1955zzd} is that SRCs were argued for as essential wavefunction features because there was \emph{no other way} to explain the cross section: 
``Consequently it follows that the usual assumptions of the shell-model theory of the nucleus, that the particles move independently in a uniform potential, cannot be other than very approximately correct.''
Each example was analyzed in the Born approximation, which gave a transparent interpretation.
For example, deuteron pickup: ejection by fast neutrons (of order 100\,MeV) of fast deuterons by nuclei.
In the picture of Fig.~\ref{fig:cartoon_nuclei}(a), the fast incident neutron would have too-small matrix elements with the mean-field protons to enable the observed deuterons to form; this required the picture of Fig.~\ref{fig:cartoon_nuclei}(b).
At this level of approximation, a momentum distribution with a tail like in Fig.~\ref{fig:cartoon_mom_dist} could be fit.

However, this clear picture was subsequently muddied by consideration of final state (and initial state) interactions (e.g., see Ref.~\cite{PhysRev.114.786}). 
The clean extraction of a momentum distribution was no longer clean.
Nevertheless, the picture associated with a hard Hamiltonian persisted and Brueckner led the way in developing methods to handle such interactions and also explain how independent-particle behavior could arise.
This proved to be a difficult program to carry out with precision using the expansion and resummation techniques of Brueckner theory.
Instead, the most successful demonstration that hard Hamiltonians (and the Argonne potential in particular) led to quantitative predictions of the low-lying spectra of light nuclei was made using quantum Monte Carlo methods~\cite{Pieper:2001mp,Carlson:2014vla}.
This might have seemed to seal the deal on which picture was correct.
Yet in the meantime, ``phenomenological'' approaches such as the shell model and nuclear energy density functionals  were successful in describing a wide range of data using soft interactions~\cite{preston1975structure,Ring:2005}.

An experimental advance that promised to shed light on nuclear momentum distributions was the development of electron scattering facilities that could detect knocked-out protons in coincidence with the scattered electrons.
At NIKHEF and other electron scattering facilities, these $(e,e'p)$ experiments were used to probe occupied shell-model orbitals and map out their momentum distributions~\cite{Dieperink:1990uk,Kelly:1996hd}.
The extracted shapes were consistent with orbitals having the appropriate separation energy and radius, but the overall normalization was significantly lower than predicted by an independent-particle model.
In the high-resolution picture, this was explained by the depletion of orbitals by both short-range and long-range correlations, of order 15--20\% for each.
Once again, however, this potentially clean resolution was spoiled by final-state interactions (despite efforts to use parallel or anti-parallel kinematics to suppress other reaction mechanisms)~\cite{Bianconi:1995mz}.
The picture was further muddied by purely theoretical considerations in Ref.~\cite{Furnstahl:2001xq}, which showed that field redefinitions or unitary transformations imply the picture of high-momentum SRC pairs is not unique.
This would turn out to be a foreshadowing of the RG reconciliation of hard and soft pictures.

With subsequent experimental and theoretical advances, the modern nuclear Rashomon effect grew more acute from both sides.
The problem of cleanly verifying the picture in Fig.~\ref{fig:cartoon_nuclei}(b) was apparently solved by experiments first at Brookhaven National Laboratory (BNL) and then more extensively at Jefferson Laboratory (JLab) that knocked out and detected \emph{both} members of SRC pairs (plus inclusive experiments with specially chosen kinematics).
By restricting to events with the appropriate kinematics in the final states, a series of experiments has established the key features listed in Sec.~\ref{sec:introduction}~\cite{Hen:2016kwk}.
At the same time, a picture closer to Fig.~\ref{fig:cartoon_nuclei}(a) (or at least with reduced high-momentum SRCs) has been behind explosive progress in ab initio (microscopic) calculations of nuclei, which have pushed well beyond where calculations with hard potentials have been computationally feasible~\cite{Hergert:2020bxy}.
To reconcile these pictures, we need to modify the hard potential into a soft one, while at the same time preserving the results from high-momentum-transfer experiments.
This can be accomplished by \emph{unitary} renormalization group evolution.

\section{Reconciliation by the renormalization group} \label{sec:reconciliation}

\subsection{Applying a unitary RG}

The renormalization group techniques developed by Wilson in the late 1960s and early 1970s formalized the ideas of block spinning introduced by Kadanoff, through which shorter distance scales in a system are averaged over, leaving only contributions from successively longer distance scales~\cite{PhysicsPhysiqueFizika.2.263,Wilson:1971bg,Wilson:1974mb}.
By eliminating these degrees of freedom (dofs), the underlying universal behavior was revealed, which was directly applied to explain critical phenomena (second-order phase transitions).
The other major class of RG applications from this period aimed to improve perturbation theory.
This use of RG dates from Gell-Mann and Low in 1954~\cite{GellMann:1954fq} and matured in applications to high-energy scattering in QCD~\cite{Gross:1973id,Politzer:1973fx}.
The basic idea is that a mismatch of external energy scales and those internal to loop integrals (i.e., sums over states) can generate large logarithms that modify the naive convergence of perturbation theory.
Running the RG can shift strength between couplings and loop integrals to minimize the impact of the logarithms~\cite{Weinberg:1981qq}.
The nuclear applications of the RG inherit features of both types of RG applications~\cite{Bogner:2006vp}.

In the early 1990's, independent efforts by Glazek and Wilson, who sought a Hamiltonian formalism for quantum chromodynamics (QCD)~\cite{Glazek:1993rc,Glazek:1994qc}, and  Wegner, who worked on condensed matter problems~\cite{Wegner:1994ab}, led to unitary RG evolution approaches that drove many-particle Hamiltonians to be increasingly energy diagonal.
This alleviates problems with small energy denominators.
The similarity RG (or SRG), particularly in the unitary flow equation form proposed by Wegner~\cite{Wegner:2001ab}, subsequently proved to be well suited for low-energy nuclear physics.
SRG applications built on a long history of unitary transformation methods in nuclear physics and especially the earlier introduction of RG for nuclear Hamiltonians by Bogner and Schwenk~\cite{Bogner:2001gq,Bogner:2001jn,Bogner:2003wn}.
The SRG is technically simpler (e.g., for evolving 3-body forces) and highly versatile; it can be applied in free space for interactions and operators~\cite{Tropiano:2020zwb}, and in the nuclear medium as a many-body solution method~\cite{Bogner:2009bt,Furnstahl:2013oba,Hergert:2020bxy}.

\subsection{Schematic look at factorized matrix elements}

\begin{figure*}[ptb]
     \centering
    (a)\includegraphics[width=0.25\textwidth]{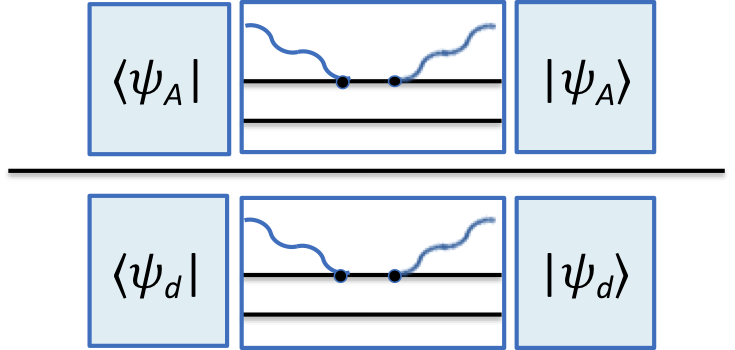}
    \quad
    (b)\!\!\!\!\!\!\!\!\!\raisebox{23pt}{{$\xrightarrow{\quad\includegraphics[width=0.15\textwidth]{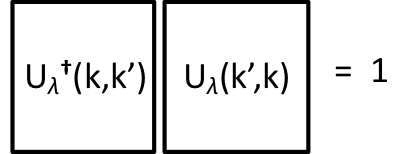}\quad}$}}
    \quad
    (c)\includegraphics[width=0.44\textwidth]{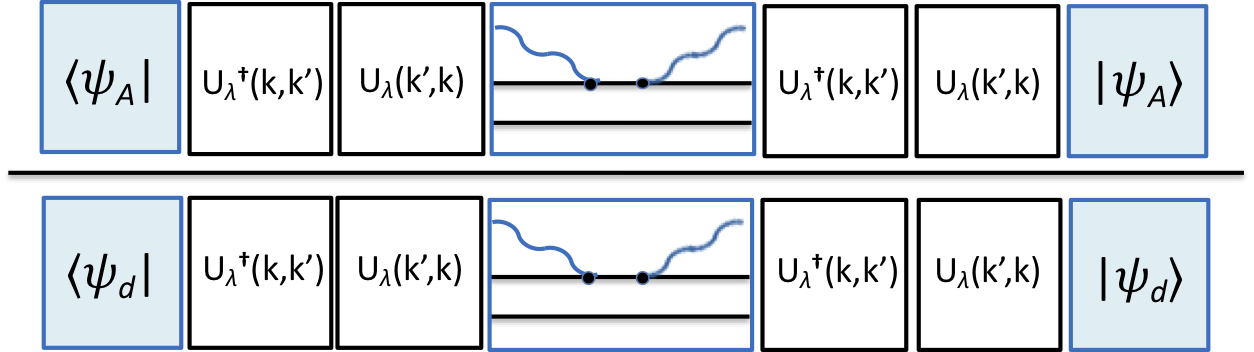}
    
    \vspace{.2in}
    \quad
    (d)\!\!\!\!\!\raisebox{18pt}{{$\xrightarrow{\quad \includegraphics[width=0.18\textwidth]{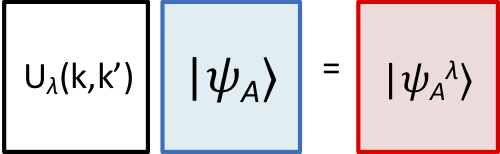}\quad}$}}
    \quad
    (e)\includegraphics[width=0.34\textwidth]{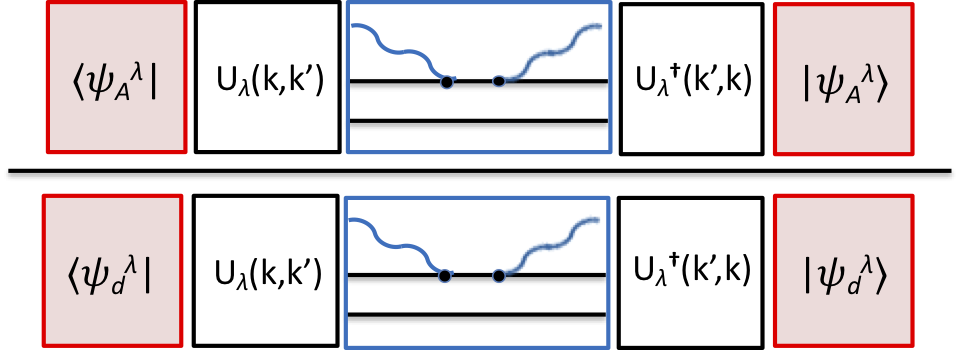}
    
    \vspace{.2in}
    \quad
    (f)\!\!\!\!\!\raisebox{18pt}{{$\xrightarrow[k < \lambda;\ k' \gg \lambda]{\quad\includegraphics[width=0.14\textwidth]{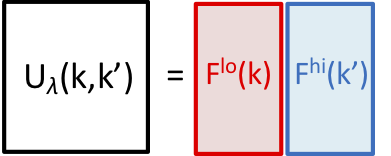}\quad}$}}
    \quad
    (g)\includegraphics[width=0.35\textwidth]{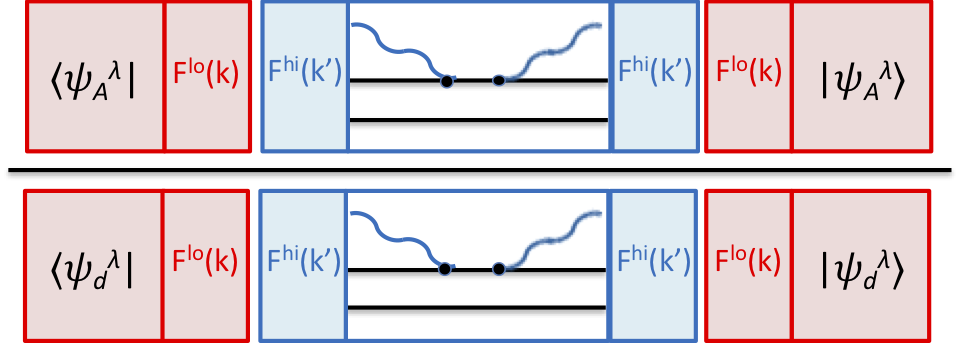}
    
    \vspace{.2in}
    \quad\raisebox{32pt}{$\xrightarrow{\hspace*{.3in}}$}\quad
    (h)\includegraphics[width=0.37\textwidth]{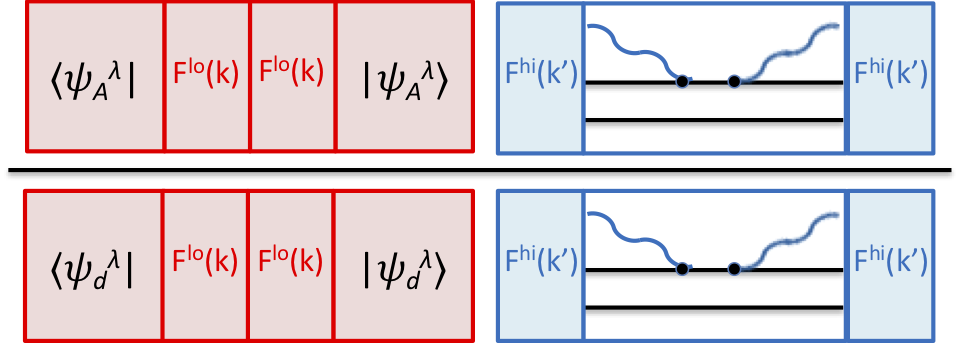}
    \quad\raisebox{32pt}{$\xrightarrow{\hspace*{.3in}}$}\quad
    (i)\includegraphics[width=0.19\textwidth]{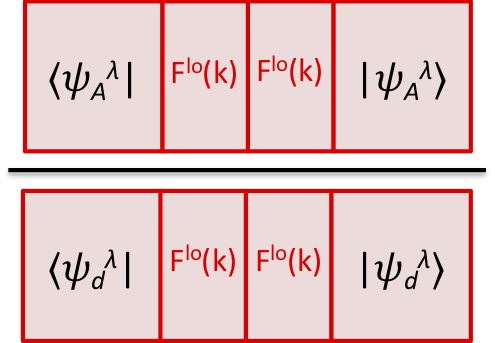}

    \caption{Schematics of what happens with high-momentum operators and low resolution states. (See the text for details.)
    }
    \label{fig:schematic_SRC_physics}

\end{figure*}

The technical aspects of the SRG are well documented in the literature~\cite{Bogner:2009bt,Furnstahl:2013oba,Roth:2013fqa,Binder:2013xaa,Hergert:2016iju}.
For our purposes we do not need to know how the unitary transformations are generated, but only their actions and characteristics.
Figure~\ref{fig:schematic_SRC_physics} provides a schematic guide to how a ratio of matrix elements of high-momentum operators is transformed from high RG resolution to low RG resolution.
For simplicity we take ground-state matrix elements and a ratio between nucleus with $A$ nucleons to the deuteron.
This setup is directly related to the experiments for inclusive cross section ratios~\cite{Egiyan:2003vg,Egiyan:2005hs,Fomin:2011ng,Ye:2017mvo,Nguyen:2020mgo}, but we can interpret the details more generally.
In Sec.~\ref{sec:low_resolution} we provide a second-quantized formulation that validates the developments sketched here.

The SRG parameter $\lambda$ is taken to be of order $\kF$ or slightly higher; it provides a dividing scale between mean-field and high-momentum physics as in Fig.~\ref{fig:cartoon_mom_dist} and it serves as the resolution scale (because only momenta less than $\lambda$ are included in the low-energy wave functions).
Details of the subplots in Fig.~\ref{fig:schematic_SRC_physics}:
\begin{enumerate}
    \renewcommand{\theenumi}{\alph{enumi}}
    
    \item The ground-state bra and ket are coded blue to indicate they are calculated using an unevolved hard interaction (e.g., AV18 plus three-body for definiteness). 
    There are both high-momentum and low-momentum contributions to $|\psi_A\rangle$ and $|\psi_d\rangle$.
    The operator is also blue, which signifies there are \emph{only} contributions from high momenta $q > \lambda$.
    For example, it could be a pair or single-nucleon momentum distribution evaluated at $q$.
    
    \item The unitary transformation at $\lambda$ is denoted $U_\lambda$. 
    Here we write it as a momentum matrix. 
    The combination $U_\lambda^\dagger U_\lambda$ is the identity matrix, so it can be inserted anywhere in our expression.
    
    \item We insert $U_\lambda^\dagger U_\lambda$ between the operator and wave function vectors with the plan of acting with the transformations on each.
    The ratio of matrix elements is unchanged.
    
    \item We designate the soft evolved ground-state matrix element with $\lambda$ and code it red. 
    Acting with $U_\lambda$ on the original wave function yields the same result as evolving the Hamiltonian and then diagonalizing to extract the ground state.
    
    \item It is evident here that to maintain the same matrix element that the operator must also be evolved by $U_\lambda O U_\lambda^\dagger$.
    
    \item When the unitary transformation is sandwiched between a part that is purely low momentum ($k < \lambda$) and a part that is purely high momentum ($q > \lambda$), then it approximately factorizes into disjoint pieces.
    This factorization is derived in Refs.~\cite{Anderson:2010aq} and \cite{Bogner:2012zm}. 
    The low-momentum part $F^{\rm lo}(k)$ is only weakly dependent on momentum.
    The factorization approximation is typically good at the 10--15\% level and corrections are calculable. 
    The leading factorization explains the dominant behavior but if greater precision is desired one can always simply use the full unitary transformation.
    
    \item Upon substitution of the factorized unitary transformation, we see that the numerator and denominator have individually separated into a purely high-momentum part that carries the full dependence on the momentum of the operator but is state independent, and a purely low-momentum part that is independent of the operator but depends on the nucleus.
    
    \item For convenience we have simply rearranged the factorized parts of the matrix elements.
    This is the leading term in an operator product expansion.
    If we focus on the red (high momentum) parts, we immediately obtain the universal (i.e., state-independent) behavior of \emph{any} high-momentum operator. Note that this applies to low-lying excited states as well.
    With the choice of a momentum distribution this is also the embodiment of 
    the GCF in Eq.~\eqref{eq:GCF_momentum} (or Eq.~\eqref{eq:GCF_coordinate} if we work in coordinate space instead).
    
    \item
    For the case of the same operator in numerator and denominator, they cancel, leaving a purely low-momentum ratio that turns out to be scale and scheme independent (to leading order).
    This is the GCF contact ratio, which is the type of ratio that will dominate the ratio of inclusive cross sections.
    Note that it is a ``mean-field'' quantity, i.e., it only depends on the soft ground-state wave function.
\end{enumerate}
Thus the GCF physics naturally emerges from a low-resolution perspective but in a form that is systematically improvable and more easily generalized.

\begin{figure}[tbh]
  \centering
  \includegraphics[width=0.7\columnwidth]{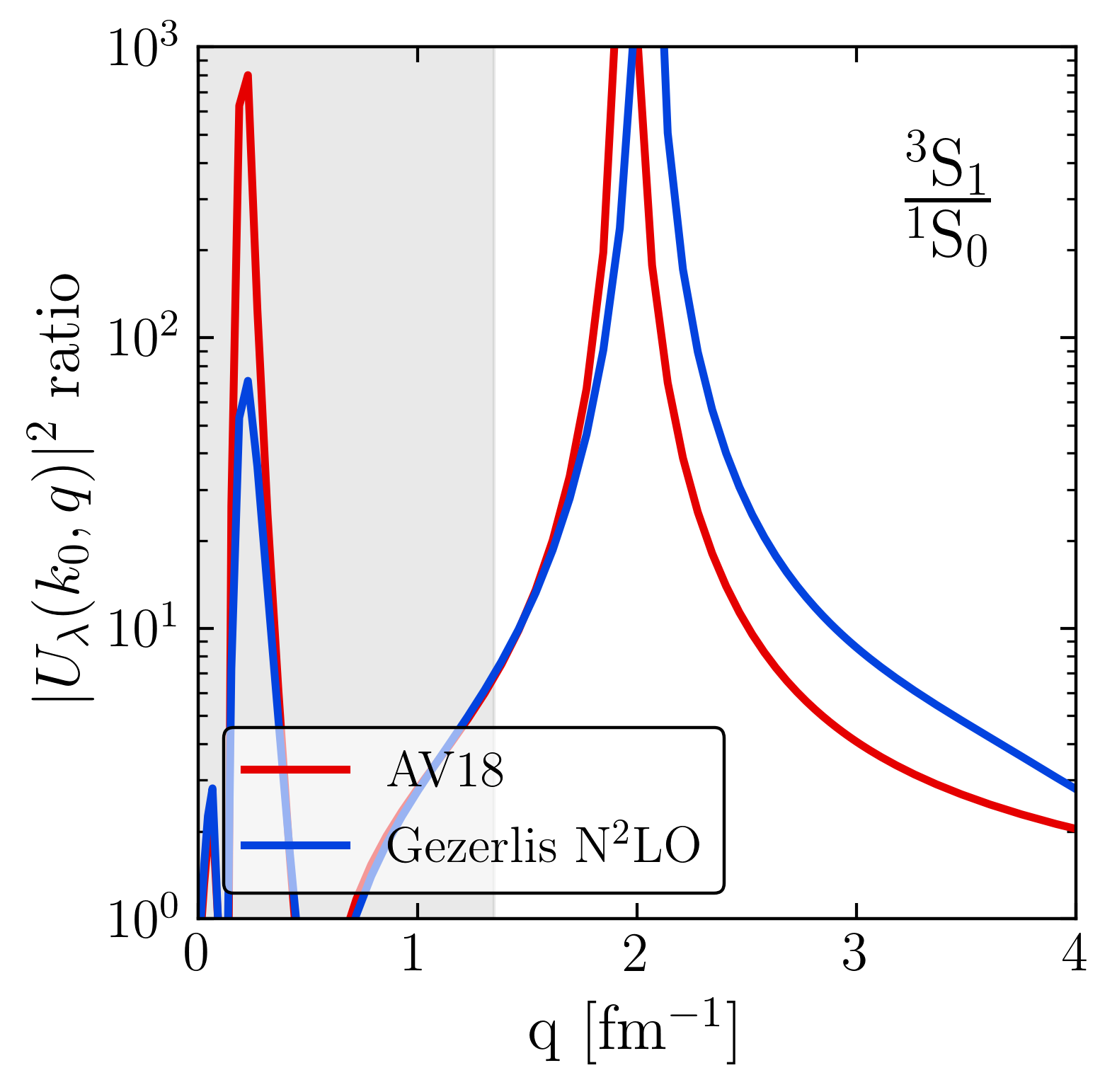}
  \caption{Ratio of $|U(k_0,q)U^\dagger(q,k_0)|$ with fixed $k_0 = 0.1\,\fmi$ for the coupled $\tripletS$ to $\singletS$ channels.
  Results are for the AV18~\cite{Wiringa:1994wb} and Gezerlis \NNLO\ (1.0\,fm)~\cite{Gezerlis:2014zia} potentials.}
  \label{fig:ratio_1p35_kvnn_6_222}
\end{figure}

Returning to (e) and (g) for the specific case of the pair distribution at high momentum, we see that it is sufficient to look at the unitary transformation of the high-resolution operator in a two- or three-body space, independent of any nucleus or state.
The two-body unitary transformation can be expanded in the NN channels without reference to a particular state (see Sec.~\ref{sec:low_resolution} for details).
In Fig.~\ref{fig:ratio_1p35_kvnn_6_222} the ratio of
$|U(k_0,q)U^\dagger(q,k_0)|$ for the coupled $\tripletS$ to $\singletS$ channels  is plotted as a function of $q$ with fixed $k_0 = 0.1\,\fmi$.
This ratio directly relates to the fraction of knocked-out pairs that are $np$ versus $pp$ or $nn$.
We see the dominance of $\tripletS$ to $\singletS$ in the region around 2\,\fmi (400\,MeV), where the tensor force is strong and the $\singletS$ potential has a node, decaying to a combinatoric fraction at high momentum.
The result is insensitive to details as the dependence on $k_0$ is very weak for $q > \lambda$.
We also see that similar results are obtained for the local chiral EFT potential from Gezerlis et al.~\cite{Gezerlis:2014zia} (this is true for any other chiral potential~\cite{Tropiano:2021prep}).

\begin{figure}[tbh]
  \centering
  \includegraphics[width=0.9\columnwidth]{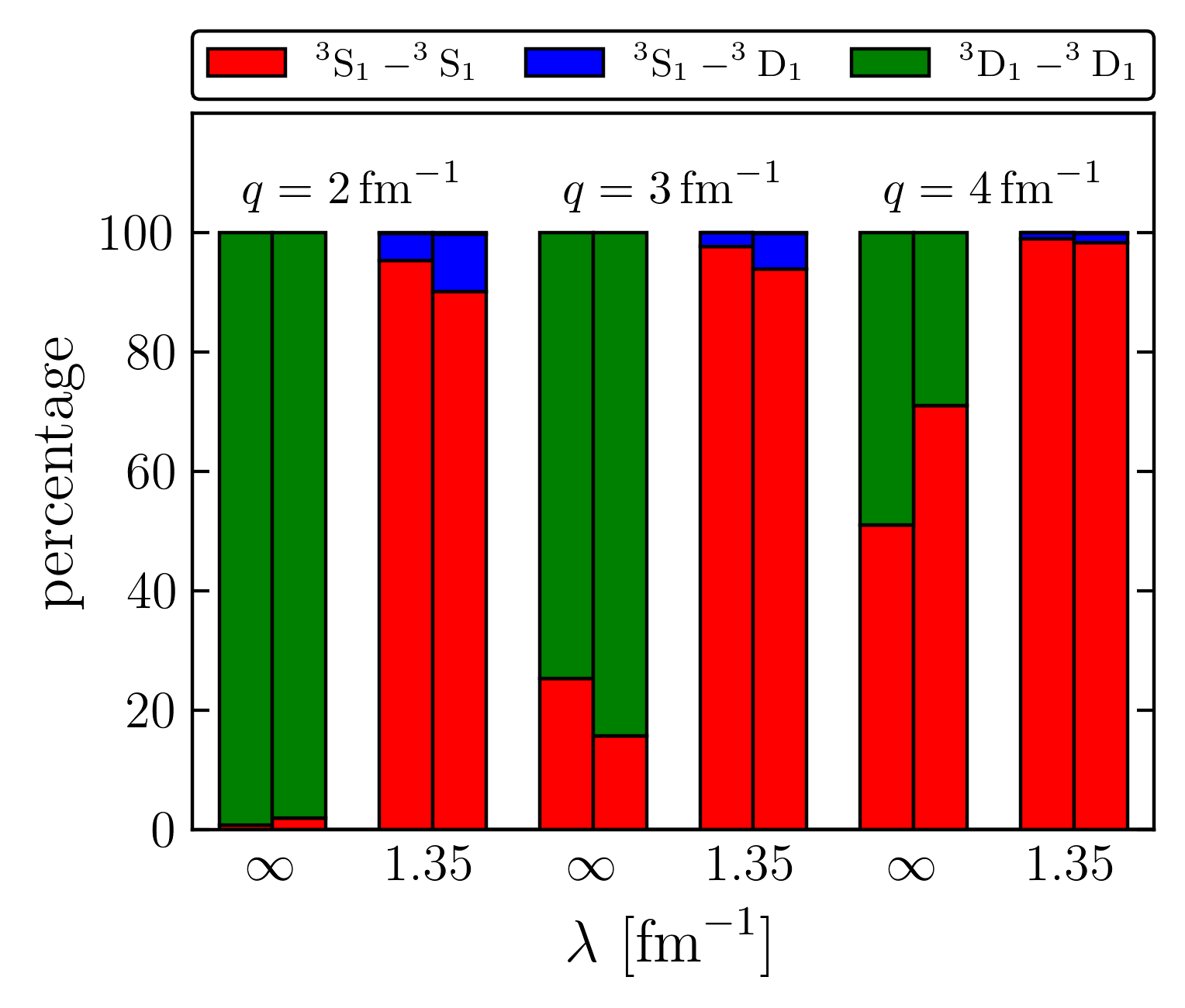}
  \caption{Percentage contributions from each channel to the matrix element
  of the momentum distribution in the deuteron with $q=2$, $3$, and $4\,\fmi$ for AV18 (left bar at each $\lambda$ value) and Gezerlis \NNLO\ (right bar) potentials. We compare unevolved and SRG-evolved (both wave function and operator, so the net matrix element is unchanged) results where $\lambda=1.35\,\fmi$.}
  \label{fig:barchart_deuteron}
\end{figure}


One might have expected that the dominance of the tensor means that the operator will be sensitive to $^3\mbox{D}_1$ admixtures in the ground state even when evolved.
This is not the case, as illustrated in Fig.~\ref{fig:barchart_deuteron}. 
The high-resolution results ($\lambda = \infty$) do come dominantly from the $\tripletD$--$\tripletD$ part for momenta $q$ where the tensor force dominates while it is more evenly split with $\tripletS$--$\tripletS$ (in a scheme/scale dependent way) at higher momenta.
But the low-resolution results ($\lambda = 1.35\,\fmi$ here) are always heavily dominated by the $\tripletS$--$\tripletS$ part.
Note that the same pattern holds for both the phenomenological AV18 potential and the Gezerlis \NNLO\ chiral EFT potential.
This S-wave dominance implies that simple ground-state wave functions with local density approximations (LDAs) will work quite well at low RG resolution, as verified in the next section.

\section{Calculating at low RG resolution} \label{sec:low_resolution}

In this section we use second quantization to more precisely characterize the SRC physics and show how simplified calculations (e.g., using LDAs) are meaningful.
The details of how factorization works in second-quantized form have already been worked out in Ref.~\cite{Bogner:2012zm}, which we build on for the current discussion.
The results presented here are only a sampling of what is possible, using the crudest approximation.
Nevertheless, they suffice to show that low-RG resolution calculations can reproduce all the features of SRC phenomenology listed in Sec.~\ref{sec:introduction}.
In future work we will explore more accurate approximations and calibrate the theoretical uncertainties~\cite{Tropiano:2021prep}.

\subsection{Second-quantized unitary transformations}
\label{subsec:second-quantization}

The SRG unitary transformation at flow parameter $\lambda$ takes the schematic second-quantized form:
\begin{align} 
    \Uhat_\lambda = \Ihat 
     &+ \sum  \delta U^{(2)}_\lambda \,
    \akdag\akdag aa
    \notag\\ 
    \null &+ \sum \delta U^{(3)}_\lambda\akdag\akdag\akdag aaa
     + [\mbox{4-body}] + \cdots
    \;,  \label{eq:Uschematic} 
\end{align}
where we have suppressed the single-particle indices and combinatoric factors.
There is no one-body term, which would have only a sum over $\akdag a$ operators.
The leading approximations illustrated in Fig.~\ref{fig:schematic_SRC_physics} can be formulated in second-quantization by always consistently truncating the operator products such as $\Uhatlam \widehat O \Uhatlamdag$ at the (vacuum) two-body level. 
In practice this is cleanly carried out by applying Wick's theorem in operator form.

The two-body term in \eqref{eq:Uschematic} in a plane-wave single-particle basis is
\begin{align} 
    \frac{1}{4} \sum_{\kvec,\kvec',\Kvec} \!  \delUtilde^{(2)}_\lambda(\kvec,\kvec') \,
    \akdag_{\frac{1}{2}\Kvec + \kvec} \akdag_{\frac{1}{2}\Kvec - \kvec} 
    \ak_{\frac{1}{2}\Kvec - \kvec'} \ak_{\frac{1}{2}\Kvec + \kvec'} 
    ,  \label{eq:Uform} 
\end{align}
where $\kvec$ and $\kvec'$ are relative momenta in the two-body subspace, $\Kvec$ is the total momentum, and $\delUtilde^{(2)}_\lambda$ is an anti-symmetrized matrix element (i.e., $\delUtilde^{(2)}_\lambda = \delta U_\lambda^{(2)} (1 + P_{12})$, where $P_{12}$ exchanges particles 1 and 2). 
For notational compactness we suppress the spin and isospin indices on $\delUtilde^{(2)}_\lambda$;
the latter dependence will be explicit later when $\delUtilde^{(2)}_\lambda$ is decomposed into the usual NN partial-wave channels.
The form in Eq.~\eqref{eq:Uform} follows from the structure of the SRG flow equation and the properties of the two-nucleon Hamiltonian, which are inherited by $\delUtilde^{(2)}_\lambda$.
Particular features of relevance in the present context are that  the first-quantized function $\delUtilde^{(2)}_\lambda$ is completely determined in the two-body system and depends on $\kvec$ and $\kvec'$ but not $\Kvec$ from Galilean invariance, three-body contributions are subleading (more on this below), and $\delUtilde^{(2)}_\lambda(\kvec,\kvec')$ factorizes as in Fig.~\ref{fig:schematic_SRC_physics}(f) for $|\kvec| < \lambda$ and $|\kvec'|\gg \lambda$.

A simple example of how truncating at the two-body level plays out is the verification of the  unitary condition in Fig.~\ref{fig:schematic_SRC_physics}(b).
We start by multiplying Eq.~\eqref{eq:Uschematic} and its hermitian conjugate and note that even maximally contracted terms involving three-body or higher terms from $\Uhat^\dagger$ or $\Uhat$ will be three-body or more, and therefore are omitted. 
Only the fully contracted part of the $\akdag\akdag a  a\akdag\akdag a a$ term from the $\delta U$ pieces will have the leading two-body form, so the net result for the leading approximation in the plane-wave basis as in \eqref{eq:Uform} is
\begin{align}
    \Uhatlamdag \Uhatlam &= \Ihat + \frac{1}{4}\sum_{\kvec,\kvec',\Kvec}
    [\delUtilde^{(2)\dagger}_\lambda(\kvec,\kvec') + \delUtilde^{(2)}_\lambda(\kvec,\kvec')
    \notag \\
    & \qquad\qquad \null + \frac{1}{2}\sum_{\kvec''} \delUtilde^{(2)\dagger}_\lambda(\kvec,\kvec'')\delUtilde^{(2)}_\lambda(\kvec'',\kvec') ]
    \notag \\
    & \qquad\qquad \null \times
     \akdag_{\frac{1}{2}\Kvec + \kvec} \akdag_{\frac{1}{2}\Kvec - \kvec} 
    \ak_{\frac{1}{2}\Kvec - \kvec'} \ak_{\frac{1}{2}\Kvec + \kvec'} 
    \notag \\
    &= \Ihat ,
    \label{eq:UdaggerU}
\end{align}
where the coefficient function in square brackets is identically zero from the unitary condition applied to (unsymmetrized) $U^{(2)}(\kvec,\kvec') = \delta_{\kvec,\kvec'} + \delta U^{(2)}(\kvec,\kvec')$.

To find the evolved form of other operators, we simply sandwich the second-quantized operator expansion between $\Uhatlam$ and $\Uhatlamdag$, apply Wick's theorem and truncate at the two-body level.
For example, the leading approximation for the evolved pair momentum distribution $\nhatlo^{\tau\tau'}(\qvec,\Qvec)$ for isospin projections $\tau$ and $\tau'$ takes the form
\begin{widetext}
\begin{align} \label{eq:pair_mom_dist_op}
     \nhatlo(\qvec,\Qvec) &=
    \Uhatlam\nhathi(\qvec,\Qvec)\Uhatlamdag \notag \\
    &\approx \nhathi(\qvec,\Qvec) 
    + \frac14\sum_{\kvec} \delUtilde^{(2)}_\lambda(\kvec,\qvec)
    \akdag_{\frac{1}{2}\Qvec + \kvec} \akdag_{\frac{1}{2}\Qvec - \kvec} 
    \ak_{\frac{1}{2}\Qvec - \qvec} \ak_{\frac{1}{2}\Qvec + \qvec}
    + \frac14\sum_{\kvec} \delUtilde^{(2)\dagger}_\lambda(\qvec,\kvec)
    \akdag_{\frac{1}{2}\Qvec + \qvec} \akdag_{\frac{1}{2}\Qvec - \qvec} 
    \ak_{\frac{1}{2}\Qvec - \kvec} \ak_{\frac{1}{2}\Qvec + \kvec} \notag \\
    & \quad \null + \frac18 \sum_{\kvec,\kvec'} \delUtilde^{(2)}_\lambda(\kvec,\qvec)\delUtilde^{(2)\dagger}_\lambda(\qvec,\kvec')
    \akdag_{\frac{1}{2}\Qvec + \kvec} \akdag_{\frac{1}{2}\Qvec - \kvec} 
    \ak_{\frac{1}{2}\Qvec - \kvec'} \ak_{\frac{1}{2}\Qvec + \kvec'},
\end{align}
\end{widetext}
where the unevolved ($\lambda = \infty$) pair momentum distribution is
\begin{equation} \label{eq:n_infinity}
    \nhathi(\qvec,\Qvec) = 
       \frac{1}{2}\sum_{\sg,\sg'} \akdag_{\frac{1}{2}\Qvec + \qvec} \akdag_{\frac{1}{2}\Qvec - \qvec} 
    \ak_{\frac{1}{2}\Qvec - \qvec} \ak_{\frac{1}{2}\Qvec + \qvec},
\end{equation}
and we have suppressed spin and isospin indices in \eqref{eq:pair_mom_dist_op} and \eqref{eq:n_infinity}.
If Eq.~\eqref{eq:pair_mom_dist_op} is summed over $\qvec$ at fixed $\Qvec$, a simple change of labels in the sums manifests that by consistently truncating to the two-body operators, the full normalization comes from the unevolved distribution, with the terms depending on $\delUtilde^{(2)}_\lambda$ canceling by the unitary condition, as in \eqref{eq:UdaggerU}. 

In practice we evaluate matrix elements of $\delUtilde^{(2)}_\lambda$ in the two-body space using antisymmetrized kets expanded in partial waves~\cite{Dickhoff:2005text}:
\begin{align}
    \label{eq:pw_ket}
   & \ket{\kvec_1 \sg1 \ta1 \, \kvec_2\sg2\ta2}  = 
    \frac{1}{\sqrt{2}} \sum_{S,M_S}\sum_{L,M_L}\sum_{J,M_J}\sum_{T,M_T}\!
    \braket{\sg1 \sg2}{S M_S} \notag \\
    & \qquad\quad \null\times \braket{\ta1 \ta2}{T M_T} 
    \sqrt{\frac{2}{\pi}}Y^*_{LM_L}(\khat) \braket{L M_L S M_S}{J M_J} \notag \\
    & \qquad\quad \null \times
    [1 - (-1)^{L+S+T}]\,
    |\Kvec k (LS) J M_J T M_T) ,
\end{align}
where 
$\kvec \equiv \half(\kvec_1 - \kvec_2)$, $\Kvec \equiv \kvec_1 + \kvec_2$,
and $\sigma$ and $\tau$ denote the spin and isospin projections, respectively.
The factor of $\sqrt{2/\pi}$ comes from the completeness relation in relative momentum space $1 = \frac{2}{\pi} \int_0^{\infty} dk k^2 \ketbra{k}{k}$.
Formulas are given in Appendix~\ref{sec:formulas} for the momentum distributions in the LDA that are used in Sec.~\ref{subsec:LDA}.

\begin{figure}[t]
    \centering
    \includegraphics[width=0.8\columnwidth]{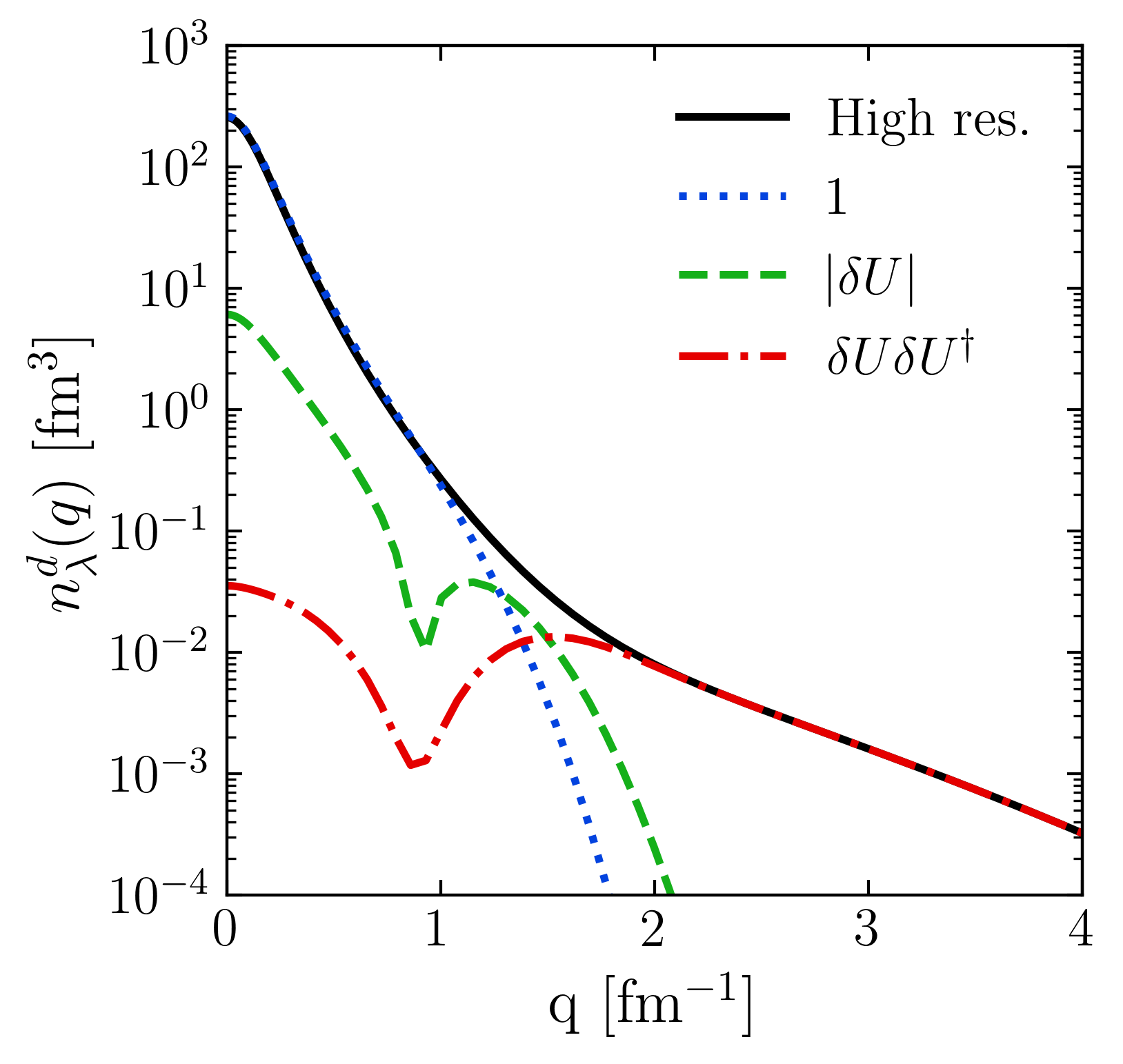}
    \caption{Contributions to the deuteron momentum distribution $n_\lambda^d(q)$ with AV18 and $\lambda=1.35$ \fmi\, from the expectation value of Eq.~\eqref{eq:pair_mom_dist_op} in the unevolved ($\lambda = \infty$, labeled ``High res.'') and evolved wave functions. The latter is split into three pieces, with the sum indistinguishable from the unevolved curve. The normalization  is $\frac{2}{\pi} \int_0^{\infty} dq\, q^2 n^d_{\lambda}(q)$ = 1.}
    \label{fig:deuteron_md_contributions}
\end{figure}

The decomposition in Eq.~\eqref{eq:pair_mom_dist_op} is exact for the deuteron. The contributions of the individual pieces are shown in Fig.~\ref{fig:deuteron_md_contributions} for the AV18 potential unevolved and evolved to $\lambda = 1.35\,\fmi$, which we take as a representative low-resolution RG scale for nuclear ground states.
The full momentum distribution at high-resolution is shown as a solid line, which is also equal to the sum of the low-resolution pieces.
The matrix element of the ``$\Ihat$'' piece from \eqref{eq:pair_mom_dist_op} (first term) has support only for low momenta; in heavier nuclei it will correspond to the mean-field region of Fig.~\ref{fig:cartoon_mom_dist}.
The terms linear in $\delUtilde$ are negative at low momentum, removing strength from this region.
(Note, the absolute value of $\delUtilde$ is shown in Fig.~\ref{fig:deuteron_md_contributions}.)
The $\delUtilde\delUtilde^{\dagger}$ piece is the sole contributor to the high-momentum region above 2\,\fmi.
The region where the ``$\Ihat$'' and linear $\delUtilde$ terms fall off while the $\delUtilde\delUtilde^{\dagger}$ term dominates is shifted to higher momenta with increasing values of $\lambda$~\cite{Tropiano:2021prep}.

When taking matrix elements of the $\delUtilde\delUtilde^{\dagger}$ term in a low-resolution wave function, $\Qvec,\kvec,\kvec'$ will all be soft, so for $|\qvec| > \kF$ the $\delUtilde$ functions will be in the factorization regime.
This will isolate the $\qvec$ dependence of the pair distribution as in Fig.~\ref{fig:schematic_SRC_physics}(h).
For the deuteron, only the \Striplet--\Dtriplet\,part of $\delUtilde$ will contribute.

How accurate will the truncation to the two-body level be for $A>2$?
Calculations of bulk quantities such as energies and radii are sensitive to three-body contributions at low resolution, but their role is amplified by cancellations of the kinetic and potential energies.
The cluster hierarchy of the potential itself is maintained in the SRG evolution.
For high-momentum distributions, we expect that the two-body contribution will dominate.
This is supported by the work of Neff et al.~\cite{Neff:2015xda} on the pair distribution in the alpha particle, which showed some $\lambda$ dependence near $2\,\fmi$ in  the dominant $S=1,T=0$ spin-isospin channel when integrating over the center-of-mass momentum $\Qvec$ (with significant dependence in the other channels) but very little dependence at $\Qvec = 0$; in all cases there is little dependence above $3\,\fmi$.
In future work we will use the RG running (i.e., the $\lambda$ dependence) to test the truncation error in our calculations~\cite{Tropiano:2021prep}.
We emphasize that corrections from three-body operators are fully accessible within our approach and that good approximations to these contributions are possible.
Finally, we note that truncation at the two-body level has much in common with the leading term of the Brueckner expansion~\cite{Brueckner:1955zzd,Tropiano:2021prep}. This has implications for SRC physics at high density, e.g., in neutron stars, where three-body physics is essential for a quantitative description.

\begin{figure*}[tbh]
	\includegraphics[clip,width=0.85\textwidth]{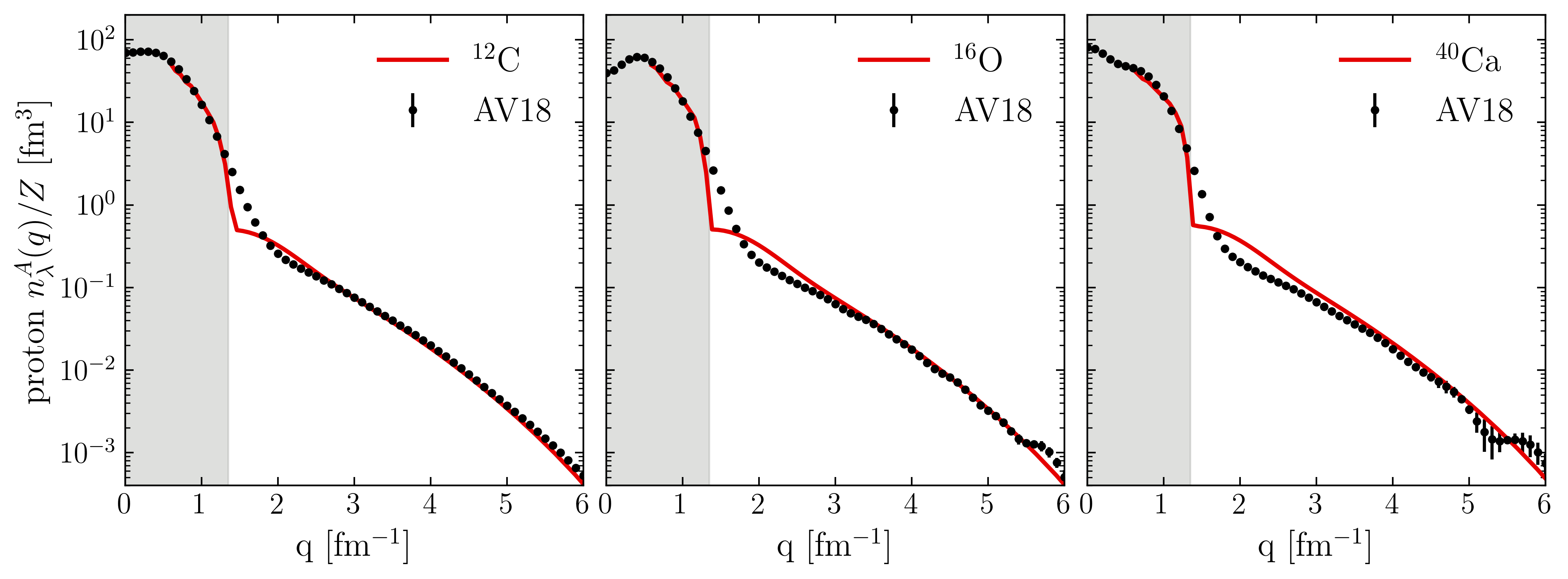}
    \caption{Proton momentum distributions for $^{12}$C, $^{16}$O, and $^{40}$Ca calculated in the LDA for $q>0.6\,\fmi$. Here we use AV18 to SRG-evolve the operator setting $\lambda=1.35$ \fmi, and divide each distribution by the proton number $Z$. The gray-shaded sections are where $q<\lambda$. Black dots correspond to AV18 quantum Monte Carlo calculations~\cite{Wiringa:single_distributions}.}
    \label{fig:proton_momentum_distributions}
\end{figure*}
%


\subsection{Local density approximation calculations}
  \label{subsec:LDA}
  
At low resolution there are various options for calculating SRC physics as manifested in the experiments described in Sec.~\ref{sec:introduction}.
By softening the Hamiltonian, the nuclear ground-state wave functions become less correlated, more amenable to many-body perturbation theory, and more universal in nature.
Furthermore, for operators evaluated at the highest momenta, the details of long-range correlations should become less important.
Indeed, the physics is focused on short distances, which suggests that an LDA should work well, particularly as the unitary transformations in the factorization region ($\delUtilde^{(2)}_\lambda(\kvec,\kvec')$ with $|\kvec| < \lambda$ and $|\kvec'|\gg \lambda$) are weakly dependent on the low momentum~\cite{Tropiano:2020zwb}.
Figures included as Supplemental Material [URL will be inserted by publisher] illustrate the factorization of the $\delUtilde\delUtilde^{\dagger}$ term in Eq.~\eqref{eq:pair_mom_dist_op}.

We illustrate how to formulate an LDA by starting with the second-quantized version of the unevolved single-particle momentum distribution for isospin projection $\tau$ from coordinate-space integrals (cf.\ Ref.~\cite{Wiringa:2013ala}):
\begin{align} \label{eq:nk_from_dm}
    n^\tau(\qvec) &= \int d\rvec \int d\rvec' \,
     \mel{\Psi_A}{\psihat^\dagger_\tau(\rvec')\psihat^{\phantom{\dagger}}_\tau(\rvec)}{\Psi_A}
      e^{-i\qvec\cdot(\rvec - \rvec')}  \notag \\
     &= \int d\Rvec \int d\svec \,
     \rhoDM^\tau(\Rvec,\svec) e^{i\qvec\cdot\svec},
\end{align}
where $\Rvec = (\rvec + \rvec')/2$, $\svec = \rvec' - \rvec$,
and $\rhoDM^\tau(\Rvec,\svec)$ is the density matrix for the $A$-body nucleus.
We implement an LDA as the leading term in a density matrix expansion (DME)~\cite{Negele:1972zp}:
\begin{align} \label{eq:dme}
    \rhoDM^\tau(\Rvec,\svec) \approx \rho^\tau(R) \rhoSL(s \kF^\tau(R)) + \cdots
\end{align}
where the Slater function is $\rhoSL(z) \equiv \frac{3}{z}j_1(z)$, the local Fermi momentum is $\kF^\tau(R) = (3\pi^2\rho^\tau(R))^{1/3}$ with $\rho^\tau$ the proton or neutron number density normalized to $Z$ or $N$, and we have applied angle averaging.
Negele and Vautherin showed this was a good approximation for not-too-large values of $s$~\cite{Negele:1972zp}.

If we substitute \eqref{eq:dme} into \eqref{eq:nk_from_dm} and integrate over $s$, we find
\begin{align}
    n^\tau(q) \approx 2 \int d\Rvec\, \theta(\kF^\tau(R) - q),
\end{align}
with the factor of 2 from the spin sum.
This is a poor approximation at very low $q$ because of the contribution to the integral from large $s$,
but its generalizations to include the SRG unitary transformations provide a quantitative
reproduction of momentum distributions at high momenta.  
All of the second-quantized terms with $\delUtilde$ will be of the form $a^\dagger a^\dagger a a$. 
These have the same structure as a Hartree-Fock energy for a non-local potential, to which we can apply the corresponding DME from Refs.~\cite{Negele:1972zp,Bogner:2008kj}.
This approximation has a single spatial integration with two $\theta$ functions featuring the local Fermi momenta.
(The explicit formulas for the low-resolution LDA momentum distributions are given in the appendix.)

We demonstrate the LDA for proton momentum distributions using the AV18 potential in Fig.~\ref{fig:proton_momentum_distributions}, for which we can compare to quantum Monte Carlo calculations~\cite{Wiringa:2013ala,Lonardoni:2017egu,Wiringa:single_distributions}.
We use proton and neutron densities generated from the SLy4 Skyrme functional~\cite{Chabanat:1997un} using the HFBRAD code~\cite{Bennaceur:2005mx}.
We expect the approximations to be valid at least for momenta above the gray-shaded regions and indeed the agreement is quite reasonable, particularly at the highest momenta.
(We exclude the predictions below $q = 0.6\,\fmi$ because of the poor approximation; better treatments will be presented in Ref.~\cite{Tropiano:2021prep}.)
It is evident that the high momentum tails are very similar across the nuclei; this is the manifestation of the universal behavior of these distributions.
There is a clear signature of the sharp cutoff caused by the $\theta$ functions in Eq.~\eqref{eq:snmd_partial_waves} at momenta near the Fermi momentum $\kF$.
We expect a smoother distribution with higher-order contributions to the DME as well as from long-range correlations.
From Table~\ref{tab:partial_wave_contributions} we see evidence that the s-waves dominate the contributions to these momentum distributions.

\begin{table}[htb]
	\caption{Percentage contributions from s-waves and selected p-waves to proton momentum distributions for $q>2\,\fmi$. We show $N_X/N$ with $N \sim \int_2^{\infty} dq\, q^2 n_{\lambda}^A(q)$, where $n_{\lambda}^A(q)$ is the proton distribution for nucleus $A$ with evolved operator at $\lambda = 1.35\,\fmi$ and $X$ denotes using only one partial wave.}
	\label{tab:partial_wave_contributions}
	\begin{ruledtabular}
		\begin{tabular}{c|ccccc}
		   Nucleus & $^{1}$S$_0$ [\%] & $^{3}$S$_1$ [\%] & $^{3}$P$_0$ [\%] & $^{1}$P$_1$ [\%] & $^{3}$P$_1$ [\%]
		   \mystrut\\
			\colrule
      			$^{12}$C & $15.5$ & $78.9$ & $1.0$ &
      			$0.8$ & $3.8$
                    \mystrut\\
      			$^{16}$O & $15.5$ & $79.1$ & $1.0$ &
      			$0.8$ & $3.6$
                    \mystrut\\
      			$^{40}$Ca & $15.4$ & $78.7$ & $1.1$ &
      			$0.9$ & $4.0$
                    \mystrut\\
      			$^{48}$Ca & $15.4$ & $78.5$ & $1.1$ &
      			$0.9$ & $4.1$
                    \mystrut\\
      			$^{56}$Fe & $15.4$ & $78.4$ & $1.1$ &
      			$0.9$ & $4.2$
                    \mystrut\\
      			$^{208}$Pb & $15.4$ & $78.4$ & $1.1$ &
      			$0.9$ & $4.2$
                    \mystrut
		\end{tabular}
  	\end{ruledtabular}
\end{table}

\begin{figure}[tbh]
    \centering
    \includegraphics[width=0.85\columnwidth]{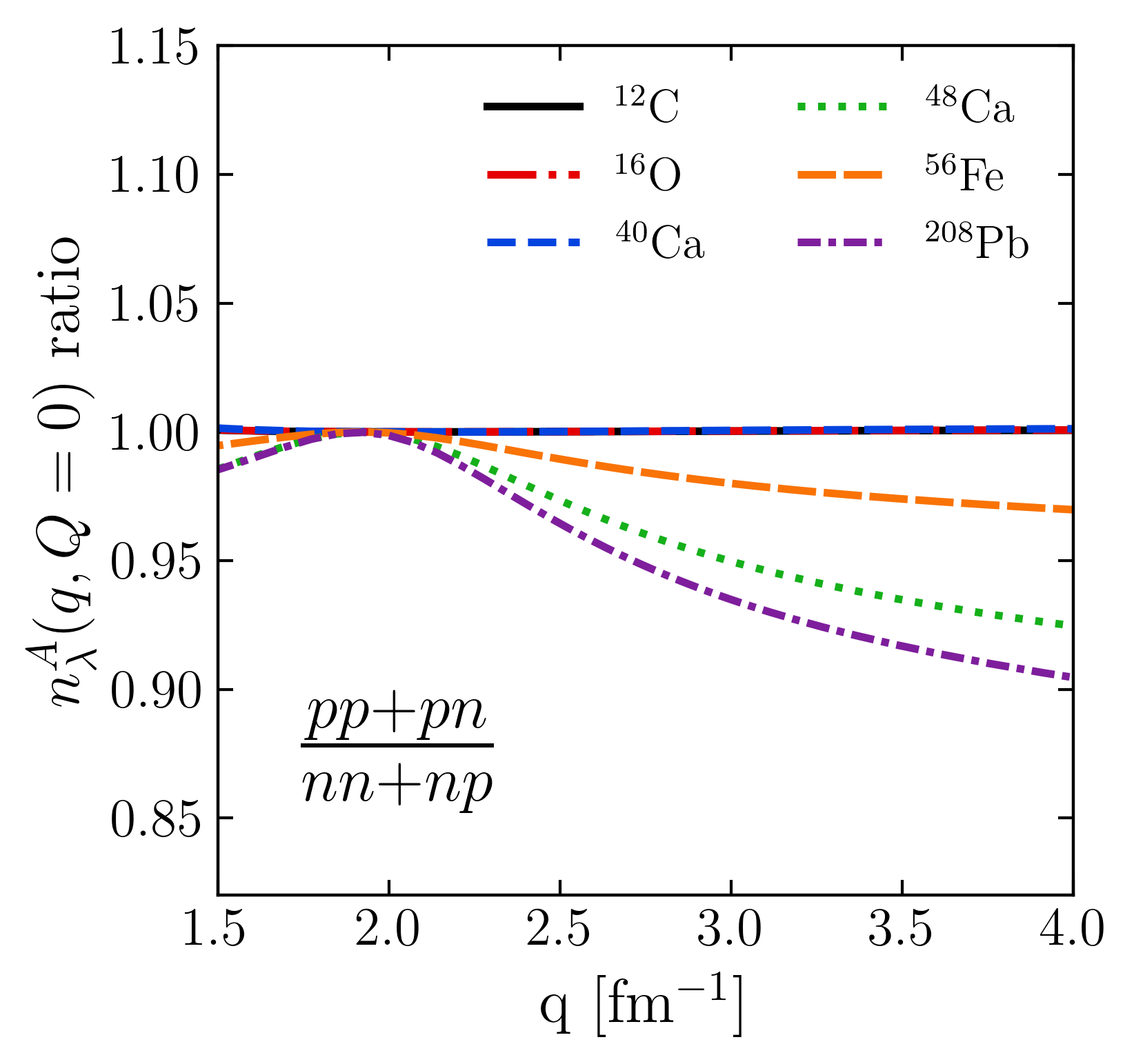}
    \caption{Ratios of pp + pn to nn + np for various nuclei calculated in the LDA with the leading two-body truncation of the pair distribution operator evaluated at $Q=0$. Here we use AV18 and set $\lambda=1.35$ \fmi.}
    \label{fig:pp_plus_pn_ratios}
\end{figure}

\begin{figure}[tbh]
    \centering
    \includegraphics[width=0.85\columnwidth]{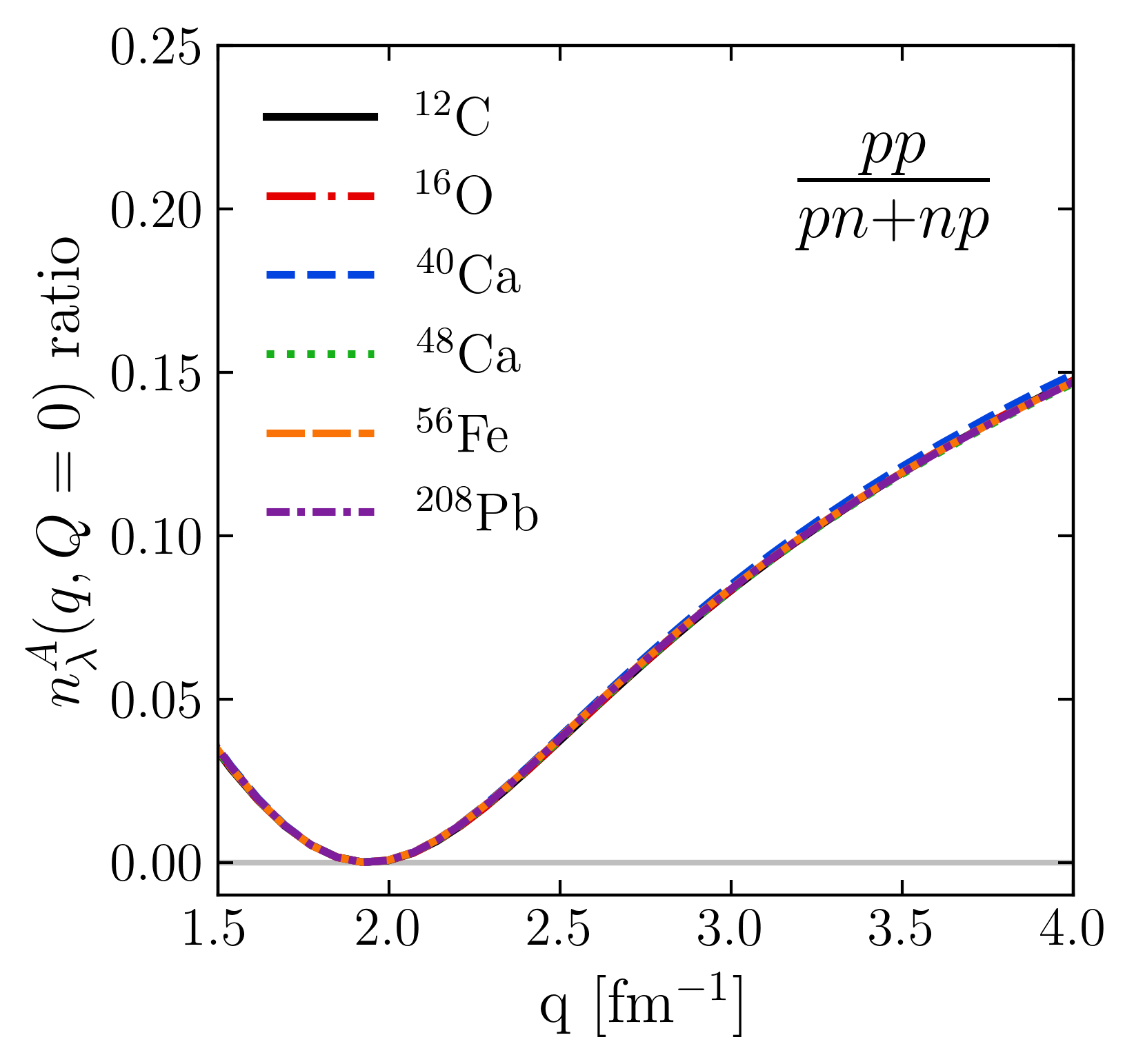}
    \caption{Ratios of proton-proton to proton-neutron distributions using AV18 and $\lambda=1.35$ \fmi.}
    \label{fig:pp_to_pn_ratios}
\end{figure}

The high-momentum tail of any momentum distribution will be scale- and scheme-dependent, i.e., dependent on the potential used. 
We can reproduce results at low-RG resolution for any initial potential, but the reaction operators must be consistently evolved.
As evident from Fig.~\ref{fig:schematic_SRC_physics}, we can cancel out the scale and scheme dependence in many cases by taking ratios.
We show examples in Figs.~\ref{fig:pp_plus_pn_ratios}, and \ref{fig:pp_to_pn_ratios} for the pair distribution at $Q=0$.
A clarifying feature of the LDA is that the theta functions defining the Fermi sea imply simple consequences for the pair distribution when $Q=0$.
In particular, the proton and neutron Fermi spheres overlap, so when $N \geq Z$ the product of the theta functions  is unity for the proton Fermi momentum below $\kF^p$, independent of $\kF^n$.

In Fig.~\ref{fig:pp_plus_pn_ratios} we show ratios of the pp + pn to nn + np pair momentum distributions at $Q=0$ as a function of $q$ for six nuclei from $^{12}$C to $^{208}$Pb.
We restrict the range of the plot to the region where SRC physics is expected to dominate.
The ratio is equal to one for $N=Z$ nuclei at all $q$, as expected from factorization as illustrated in Fig.~\ref{fig:schematic_SRC_physics} if the nucleus in numerator and denominator are the same.
Even though there are no high-momentum nucleons in the low-resolution wave function, the result is consistent with phenomenology that the proton and neutron high-momentum distributions should be about the same~\cite{Hen:2016kwk}.
In the region of np dominance near $q=2\,\fmi$ (400\,MeV), this ratio should be unity independent of $N/Z$, while there is $N/Z$ dependence away from this region because there are more nn than pp pairs.

In Fig.~\ref{fig:pp_to_pn_ratios} we show ratios of the pp to pn pair distributions as a function of $q$ for the same nuclei as in Fig.~\ref{fig:pp_plus_pn_ratios}.
The ratio dips down to essentially zero just before 2\,\fmi and then rises to about 0.15 at 4\,\fmi.
This trend is also observed in the GCF calculations~\cite{Cruz-Torres:2019fum,CLAS:2020rue} and reflects the transition from the dominant effect of the tensor force toward the scalar limit.
In the low-resolution calculations here, this dependence on $q$ is purely from the high-momentum part of the factorized unitary transformations.
The ratio shows no dependence on $N/Z$ because the pair momentum distribution at high-$q$ and $Q=0$ is restricted entirely by the proton Fermi sphere in the $\delUtilde\delUtilde^{\dagger}$ term of Eq.~\eqref{eq:pmd_partial_waves} for $N \ge Z$ nuclei.

\begin{figure}[tbh]
    \centering
\includegraphics[width=0.82\columnwidth]{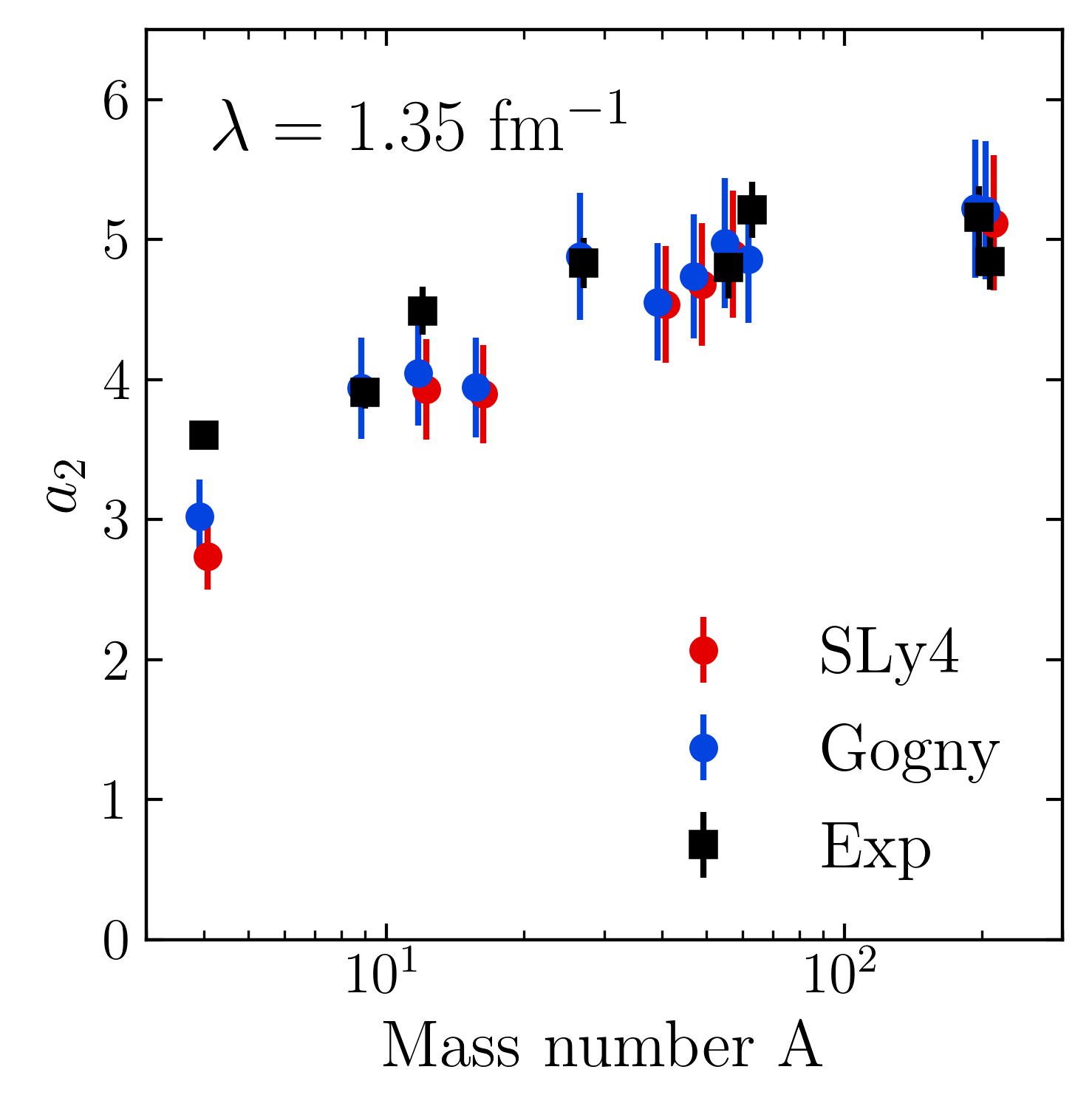}
    \caption{SRC scaling factors $a_2$ calculated using AV18 according to Eq.~\eqref{eq:a2}. See text for further details.}
    \label{fig:src_scaling_factors_a2}
\end{figure}

Finally, in Fig.~\ref{fig:src_scaling_factors_a2} we compare calculated SRC scaling factors $a_2$ using AV18 to values for nuclei extracted from inclusive cross section ratios~\cite{Fomin:2011ng,Schmookler:2019nvf}.
Following the prescription of Ref.~\cite{Ryckebusch:2019oya}, we estimate $a_2$ by integrating single-nucleon probability distributions (phase-space weighted $n_{\lambda}^A(q)$, summing proton and neutron distributions) over a high-momentum range:
\begin{align}
    \label{eq:a2}
    a_2(A) = \lim_{q\to\infty} \frac{P^A(q)}{P^d(q)} \approx \frac{2}{A} \frac{\int_{\Delta q^{\rm high}} dq\, q^2 n_{\lambda}^A(q)}{\int_{\Delta q^{\rm high}} dq\, q^2 n_{\lambda}^d(q)}.
\end{align}
We apply the LDA to both the numerator and denominator, which may help to cancel systematic errors.
Results using densities from the SLy4 Skyrme functional~\cite{Chabanat:1997un} are shown in red and results using densities from the Gogny functional~\cite{Decharge:1979fa} in blue.
The ``error bars'' (which are not statistical uncertainty intervals) are established by using integration ranges $\Delta q^{\rm high} = 2$--$\infty$ \fmi and  $\Delta q^{\rm high} = 3.8$--$4.5$ \fmi~\cite{Ryckebusch:2019oya}.
The agreement with experiment is consistent with these error bars.

The magnitude of $a_2$ and the trend with $A$ can be roughly understood at low-RG resolution from a mean-field treatment of the leading operator product expansion for the integrands in \eqref{eq:a2}~\cite{Anderson:2010aq,Chen:2016bde,Lynn:2019vwp}. This means a factorization approximation to $\delta U(k,q)$ (including constant $k$ dependence) and the dominance of the $\tripletS$ contribution.
When applied to the third term of Eq.~\eqref{eq:snmd_partial_waves}, which is the only contribution to the $a_2$ integrals,
the short-distance $q$ dependence cancels in the $a_2$ ratio and we are left with integrals of the $\theta$ functions over total momentum $K$ and relative momentum $k$. 
The integrals decouple after switching to single-particle momenta,  yielding a product of the proton and neutron densities at $R$. Thus $a_2$ is simply given by the ratio of $(1/A)\int d^3R\, \rho_p(R)\rho_n(R)$ factors, as would be expected from the schematic treatment in Fig.~\ref{fig:schematic_SRC_physics}. 
This ratio accounts for about 80\% of $a_2$ and the variation of the integral with $A$ tracks the $A$ dependence from $^4$He to $^{208}$Pb quite closely.
The remaining 20\% comes primarily from the $\singletS$ contribution, with the relative weighting of $\singletS$ and $\tripletS$ dependent on the relative $q$ dependence in the $\delta U$ matrix elements seen in Fig.~\ref{fig:ratio_1p35_kvnn_6_222}.


\subsection{Low-resolution take on SRC physics}

Here we revisit the key features of high-resolution SRC phenomenology from Sec.~\ref{sec:introduction}, now understood from low resolution via Fig.~\ref{fig:schematic_SRC_physics} and through our explicit LDA calculations from Sec.~\ref{subsec:LDA}.
\begin{enumerate}
   \item \textbf{Universal high-momentum nucleon distributions.}
  The entire sequence in Fig.~\ref{fig:schematic_SRC_physics} illustrates the outcome of taking ratios of inclusive cross sections.
  Figure~\ref{fig:schematic_SRC_physics}(h) in particular is the embodiment of universal distributions, with the dependence on the nucleus factorized from the dependence on the high momenta.
  Explicit calculations of the momentum densities manifest the factorization as seen in Fig.~\ref{fig:proton_momentum_distributions}, where each nucleus exhibits the same $q$ dependence.
  The kinematic thresholds selected for the inclusive experiments are those necessary for this clean factorization to hold.
  For SRG parameter $\lambda$ comparable to $\kF$, this corresponds to the conditions for factorization as well as complete dominance of the two-body current.
  In taking ratios, the high-momentum part cancels out, so the leading dependence of the plateau heights in the cross-section ratio, namely $a_2$, is a mean-field quantity~\cite{Anderson:2010aq,Cruz-Torres:2019fum}, which is well reproduced in our approximations (see Fig.~\ref{fig:src_scaling_factors_a2}).
  The conditions here hold for any high-momentum current.

   \item \textbf{Kinematics of the knocked-out nucleons.}

   If we look at Fig.~\ref{fig:schematic_SRC_physics}(e), we see that the external high-momentum current operator is connected to the soft, evolved low-energy wave function by the operator $\Uhat^\dagger$.
   That wave function has mostly Fermi sea nucleons and nearby admixtures.
   When $\Uhat^\dagger$ acts, we can read off from Eq.~\eqref{eq:Uschematic} that connecting to a high-momentum nucleon must come from the $\delta U_\lambda^\dagger$ part.
   Since this part starts with a two-body piece with the same pair center-of-mass momentum $\Kvec$ as for the states annihilated from the Fermi sea or near the Fermi surface, the high-momentum states must be a pair with center-of-mass momentum distribution the same as the Fermi sea.
   Furthermore, the contributions are dominated by relative s-waves (see Table~\ref{tab:partial_wave_contributions}).
   Thus this basic part of SRC phenomenology~\cite{Cohen:2018gzh} follows immediately.
   
   We can address the role of the omitted three-body terms either through direct evolution, which is nontrivial but well-established technology for SRG evolution~\cite{Jurgenson:2009qs}, or indirectly by considering the dependence on the SRG flow parameter $\lambda$. Weak dependence indicates small effects of three-body (and higher-body) operators.
   The work of Neff, Feldmeier, and Horiuchi~\cite{Neff:2015xda} used the indirect method for the SRG-evolved nucleon pair distributions for $A=3$ and $A=4$. 
   For example, they look at pair densities with SRG two-body-only unitary transformations in helium-4 and identify where there is $\lambda$ dependence. 
    There is almost no dependence for $K=0$ pairs, but with larger center-of-mass momentum it is significant, meaning that three-body contributions cannot be neglected.
    When integrated over $K$ there is slight $\lambda$ dependence near $2\,\fmi$ in the dominant $S=1,\,T=0$ channel and strong dependence in the other channels. 
    In all cases the $\lambda$ dependence is small above $3\,\fmi$.
   The $\lambda$ dependence for the calculations presented here will be explored in detail in future work~\cite{Tropiano:2021prep}.

  \item \textbf{Ratio of $np$ to $pp$ knocked-out pairs for intermediate relative momentum (300--500 MeV).}

    Tensor dominance follows directly from calculations of the unitary transformation operator in the two-body subspace.
    Figure~\ref{fig:ratio_1p35_kvnn_6_222} shows the ratio of unitary transformations in the dominant s-waves. 
    The ratio of triplet to singlet shows the peak near $q = 2\,\fmi$ for both potentials, associated with a zero in the \singletS\ channel and the enhanced tensor contribution in the \tripletS\ channel.
    Figure~\ref{fig:barchart_deuteron}  show that despite the tensor dominance, at low-RG resolution the coupling to the wave function is through the s-waves, not tensor correlations in the wave functions.
    The consequences for experimental ratios are illustrated in Figs.~\ref{fig:pp_plus_pn_ratios} and \ref{fig:pp_to_pn_ratios}, which are consistent with SRC phenomenology.

   \item \textbf{Ratio of knock-out cross sections from neutron-rich nuclei compared to $N=Z$ nuclei.}

    In the region of $np$ dominance, the pairs coupled from the Fermi sea are almost all $np$ pairs, so the high-momentum proton and neutron distributions are the same, independent of the $N/Z$ ratio in the mean-field part.
    This is seen in Fig.~\ref{fig:pp_plus_pn_ratios} and is an immediate consequence of the $\delta U$ part being dominated by the $\tripletS$ channel and this term in $\Uhat$ creating pairs. 
    So whatever low-energy wave function it hits, it will always ``kick above the Fermi sea'' an equal number of neutrons and protons.

  \item \textbf{Transition from $np$ dominance of SRC pairs to ratios expected from scalar counting.}
   This transition is manifested in Figs.~\ref{fig:ratio_1p35_kvnn_6_222} and \ref{fig:pp_to_pn_ratios}.
   In the low-RG resolution formalism, it is simply a consequence of the two-body physics that is encapsulated in the two-body unitary transformations by the SRG evolution.
   This physics is well-known from NN scattering.

\end{enumerate}

\textbf{The generalized contact formalism.}
As already noted, the GCF phenomenology is built on
a factorization ansatz for the many-body wave function  when a pair of nucleons has small relative distance or high relative momenta.
This leads to the pair distributions in coordinate and momentum space taking the forms in Eqs.~\eqref{eq:GCF_coordinate} and \eqref{eq:GCF_momentum}, respectively.

The parallel formulation at low-RG resolution is that the dependence on $r$ or $q$ in the wave function becomes the corresponding part of the factorized unitary transformation.
As such, it is the leading term in an operator product expansion~\cite{Anderson:2010aq,Bogner:2012zm,Chen:2016bde,Tropiano:2020zwb}.
A key advantage of the RG formulation is that systematic corrections are well defined.

\textbf{The low-order correlation operator approximation (LCA)}.
The LCA methodology can be used to compute observables for nucleon knockout reactions that are dominated by SRCs.
The high-resolution ground-state wave function $|\Psi_A\rangle$ or nucleus $A$  is related to a simple wave function $|\Phi_A\rangle$ (a Slater determinant in practice) through a correlation operator
$\calG$:
\begin{align}
    |\Psi_A\rangle = \frac{1}{\sqrt{\langle \Phi_A | \calG^\dagger \calG | \Phi_A\rangle}}
    \calG | \Phi_A \rangle
    \;.
\end{align}
The LCA consists in approximating $\calG$ by central (Jastrow), tensor, and spin-isospin SRC correlations with terms corresponding to two-body operators.
The dominant contribution to the SRC part of the nuclear momentum distribution takes the form~\cite{Ryckebusch:2019oya}
\begin{align} \label{eq:LCA}
    n^A_{\rm\scriptscriptstyle SRC}(\qvec) &\sim 
    \sum_{NN'\in\{p,n\}} \sum_{\alpha\beta} \sum_{\Kvec,\kvec,\kvec'}
    \calG^\dagger_{12}(\frac{\Kvec}{2}+\kvec-\qvec) \notag \\
    & \qquad\qquad\qquad\qquad\qquad \null \times
    \calG_{12}(\frac{\Kvec}{2}+\kvec'-\qvec) \notag \\
    & \null \times \langle N\alpha, N'\beta |
    \akdag_{\frac{\Kvec}{2}+\kvec}
    \akdag_{\frac{\Kvec}{2}-\kvec}
    a^{\phantom{\dagger}}_{\frac{\Kvec}{2}+\kvec'}
    a^{\phantom{\dagger}}_{\frac{\Kvec}{2}-\kvec'}
    | N\alpha, N'\beta \rangle
    ,
\end{align}
where the $\alpha,\beta$ sum runs over  occupied single-particle states
in the Slater determinant $|\Phi_A\rangle$.
The parallels to our low-RG resolution formulation are apparent, both using a simple many-body wave function and with the correlation function taking the place of our two-body unitary transformations.
Note that these functions 
are not unitary and they are constructed by matching to ab initio results rather than being constructed directly.
It will be interesting to make more detailed comparisons.

\section{Discussion: Takeaway points} \label{sec:implications}


In this section we summarize some of the takeaway points from an RG-based perspective on SRC physics in the form of answers to frequently asked questions.

\begin{itemize}
    \item \textbf{What is short-range-correlation (SRC) physics?}
    SRC physics is short-distance physics manifesting at high RG resolution as high-relative-momentum nucleon pairs, while such pairs are suppressed at low RG resolution.
    The SRC momentum scale starts where the tensor force dominates the NN interaction (around $2\,\fmi = 400$\,MeV or a bit lower).
    The SRC pairs have center-of-mass momentum distributions with a width comparable to the Fermi momentum.

    \item \textbf{Where does the short-range physics of SRC pairs appear at low RG resolution?}
    With decreasing RG resolution, this physics will shift from wave functions (structure) to interaction operators (potential/reactions) with smooth momentum dependence. 
    This was demonstrated explicitly and quantitatively for deuteron electrodisintegration~\cite{More:2015tpa,More:2017syr}.
    The characteristics of this SRC physics can be identified from the unitary transformation operator evaluated in a few-body space (it is dominantly two-body).

   \item\textbf{Which is the correct picture of nuclei, with hard or soft potentials?}
   The RG explains that both hard and soft pictures are correct descriptions of nature (actually there is a continuum of pictures!), if one treats structure and reactions consistently (i.e., at the same RG resolution).
   
   \item\textbf{What is the best choice of RG resolution scale?} 
   Weinberg’s 3rd Law of Progress in Theoretical Physics~\cite{Weinberg:1981qq} states ``You can use any degrees of freedom you like to describe a physical system, but if you choose the wrong ones you’ll be sorry!'' 
   For applications to hard scattering in QCD, the resolution scale is chosen to be of order the characteristic four-momentum transfer so that the reaction mechanism can be calculated in perturbation theory and factorization is well established.
   But the best choice of scale may not be so clear for analyzing SRC experiments because of the trade-offs.
   
   \item\textbf{Is SRC physics missing from low-RG resolution descriptions of nuclei?}
   It is not missing. The Hamiltonians at low resolution are constructed to match energies or scattering observables for low-energy bound states. 
   This can be seen  explicitly through RG evolution (e.g., with the SRG), where short-distance modification of wave functions in coordinate space due to repulsive core or tensor interactions, or the related high-momentum tail of momentum-space wave functions, is smoothly suppressed with the lowering of the RG scale and the potential shifts by weakly momentum-dependent pieces.
   If appropriate (i.e., consistent) operators are omitted, then SRC physics \emph{will} be missed, but it can always be accommodated in a consistent calculation.

   \item\textbf{What are the advantages of a high-RG-resolution description of nuclei?}
   A high-RG description enables models of short-distance physics that use resolved degrees of freedom. 
   This can mean that the dominant reaction mechanism is particularly simple (e.g., one-body currents). 
  The high-RG-resolution GCF phenomenology provides a reasonable, if not yet high-precision, description of SRC experiments to date.
   It is also the starting point for evolution to low resolution. 

   \item\textbf{What are the advantages of a low-RG-resolution description of nuclei?}
   The nuclear structure is increasingly perturbative at low resolution. 
   Many ab initio methods only work at lower resolutions and many-body perturbation theory becomes usable. 
   Scale separations and factorization are better exploited. 
   Universal behavior is more apparent. 
   Final-state interactions have been shown to be suppressed for deuteron electrodisintegration and this is expected to be a general phenomena based on local decoupling.
   
   \item\textbf{What does it mean for structure and reaction models to be \emph{consistent}?}
   There are several aspects to consistency. 
   One aspect is the scale and scheme. 
   If one uses a low-resolution wave function with high-resolution reaction operators, this is inconsistent and can lead to apparent quenching. 
   To maintain consistency at different scales, one can start with a consistent Hamiltonian and current operators and evolve them together with an RG approach like SRG.

   \item\textbf{How do you connect the pictures at two resolutions?}
   Using the (similarity) renormalization group! The consistent low-RG resolution operators can be evolved from high resolution (even many-body operators).   Unitary evolution is an important aspect if we want to describe high-experimental-resolution probes of low-RG-resolution nuclei.
   Or consistent operators can be fit! (The latter possibility will be explored in future work~\cite{Tropiano:2021prep}.)

  \item\textbf{How does RG resolution relate to experimental resolution?}
  RG resolution is set by the decoupling scale, which dictates what momenta are part of a low-energy wave function. 
  Note that for unitary RG evolution, like the SRG, this is a separation scale; the high-momentum components are not eliminated but appear only in wave functions of high-energy states.
  Experimental resolution is set by the kinematics of the experiment. 
  These are independent!
  The RG resolution is not a measurable quantity but rather a choice of the analysis.
  
  \item\textbf{Can high-momentum nuclear distributions be measured experimentally?}
  They can be extracted from experiment in a scale/scheme dependent manner. This is the same situation as with QCD parton distributions.
  Note that this means that high-momentum distributions using different Hamiltonians (which will manifest scale and scheme dependence) will not agree.

  \item\textbf{When you soften a Hamiltonian, do you “harden” the interaction operators?}
  No. SRG makes unitary transformations, so no physics is lost, but it is nonperturbatively “reshuffled”.
  (Note that this means that an SRG-evolved Hamiltonian will have the same limitations or virtues as the original Hamiltonian!)
  Low-energy states filter operators so only low-momentum components of the operator contribute to matrix elements. 
  This leads to numerous simplifications.
  Purely high-momentum operators (meaning operators connecting to only high-momentum physics) factorize and the dependence on high momentum is state independent.
  This is the manifestation of an operator product expansion.
  Where the tensor interaction plays a role, there is a de-emphasis of D-wave physics compared to S-wave contributions.
  Reduced final-state interactions are observed.
  The complication is from higher many-body operators, but these can be evolved in few-body spaces.

  
  
  \item\textbf{What are the implications for other knock-out reactions?}
  Analyses of intermediate energy nucleon knock-out experiments often
 mix a high-resolution reaction mechanism (e.g., eikonal model) with a low-resolution structure description (e.g., shell model)~\cite{Tostevin:2014usa,Aumann:2020tcq}.
  Such a mismatch applied to the electron-scattering SRC experiments considered here would lead to essentially zero cross section predicted theoretically because the contribution from the two-body current at low RG resolution would be entirely omitted and the one-body current has no support at high momentum.
  This implies that RG evolution of the reaction operators may be relevant for resolving systematic discrepancies between measured cross sections and theoretical predictions~\cite{Tostevin:2014usa,Aumann:2020tcq} (see also Ref.~\cite{Wylie:2021uot}).

\end{itemize}

\section{Summary} \label{sec:summary}

We have demonstrated that high-RG-resolution SRC physics is faithfully incorporated at low resolution by unitary RG evolution, with weakly correlated wave functions and simple evolved operators.
We have also confirmed that at low-resolution the consequences of SRC experiments follow directly from basic and well-established nuclear physics: the density dependence of nuclei and  characterisitcs of the nucleon-nucleon interactions, in particular the tensor force and short-distance repulsion.
Furthermore, the basic features of the SRC phenomenologies, in particular the GCF and the LCA, emerge naturally by consideration of low resolution. 
The systematic SRG framework points to how to improve them.
    
The value so far  of the SRC knock-out experiments is not new insight into the internucleon interactions or features of the many-body wavefunctions.
Indeed, we have seen that all that is needed to explain the observations are familiar features of the NN interaction from phenomenological or chiral EFT potentials well understood from pion exchange and fitting to NN scattering data.
Rather, it is the demonstration that this physics can be isolated and controlled that opens the possibilities to understand more complicated reactions.
A better understanding of how to analyze reactions is critical for rare isotope nuclear physics and extensions of the SRC experiments can be a gateway.
Combined with RG analyses, such experiments can help calibrate and test reactions cleanly and set the stage for extensions to more complicated knock-out reactions.
    
        
        
        
        
        

Planned extensions of the present investigations in the short term include~\cite{Tropiano:2021prep}:
further examining the SRG resolution ($\lambda$) dependence; improving the treatment of the ground-state wave function with DME corrections; relaxing the truncation to two-body operating, quantifying the corrections and seeking tractable approximations; and re-examining $(e,e'p)$ knock-out reactions  studied at NIKHEF and other electron scattering facilities~\cite{Dieperink:1990uk,Kelly:1996hd}.
Work on all of these areas is in progress.


\begin{acknowledgments}
We thank Nathan Parzuchowski for contributions to an earlier version of this work, Nicolas Schunck for sharing a code to calculate Gogny nucleon densities, and Wim Cosyn and Jan Ryckebusch for helpful comments on the manuscript. We thank our colleagues for many useful discussions on SRC physics over the years.
The work of AJT and RJF was supported by the National Science Foundation under Grant No.~PHY--1913069, and the NUCLEI SciDAC Collaboration under US Department of Energy MSU subcontract RC107839-OSU\@.
The work of SKB was supported by the National Science Foundation under Grant Nos. PHY-1713901 and PHY-2013047, and the U.S. Department of Energy, Office of Science, Office of Nuclear Physics under Grant No. de-sc0018083 (NUCLEI SciDAC Collaboration).
\end{acknowledgments}

\newpage

\appendix

\section{LDA formulas} \label{sec:formulas}

In this appendix we summarize the LDA formulas applied in Sec.~\ref{subsec:LDA}.
The single-nucleon momentum distribution is:
\begin{widetext}
\begin{align}
    \label{eq:snmd_partial_waves}
    & n_{\lambda}^{\tau}(q) =  \int d^3R\, \Bigl\{ 2 \theta(\kF^{\tau}-q) \notag \\
    & \quad \null + 32  \sum_{L,S,T}\!\!{\vphantom{\sum}}'\, \sum_{J}  (2J+1)
    \frac{2}{\pi}\int_0^{\infty}\! dk\, k^2  
    \melr{k (L S) J T}{\delta U}{k (L S) J T}  
        \sum_{\tau'} |\braket{\tau \tau'}{T \, \tau+\tau'}|^2
\theta(\kF^{\tau}-q) \int_{-1}^1 \frac{dx}{2}\theta(\kF^{\tau'}-\abs{\mathbf{q}-2\mathbf{k}}) \notag \\
    & \quad \null + 2  \sum_{L,L',S,T}\!\!\!\!\!{\vphantom{\sum}}'\,\,  \sum_{J}  (2J+1)  
    \Bigl(\frac{2}{\pi}\Bigr)^2 \int_0^{\infty} dk\, k^2 \int_0^{\infty} dK\, K^2 \int_{-1}^1 \frac{dy}{2} \int_{-1}^1 \frac{dz}{2}  
    \melr{k (L S) J T}{\delta U}{\abs{\mathbf{q}-\mathbf{K}/2}  (L' S) J T} 
    \notag \\
    & \qquad\qquad\qquad \null\times   
     \melr{\abs{\mathbf{q}-\mathbf{K}/2} (L' S) J T}{\delta U^{\dagger}}{k (L S) J T}  
    \sum_{\tau'} |\braket{\tau \tau'}{T \,\tau+\tau'}|^2
     \,\theta(\kF^{\tau}-\abs{\mathbf{K}/2 + \mathbf{k}}) \,\theta(\kF^{\tau'}-\abs{\mathbf{K}/2 - \mathbf{k}})
     \Bigr\},
 \end{align}
%
where the local Fermi momentum is $\kF^\tau(R) = (3\pi^2\rho^\tau(R))^{1/3}$ with $\rho^\tau$ the proton or neutron number density normalized to $Z$ or $N$.
For the angle averaging we have defined 
\begin{align}
  \abs{\mathbf{q}-2\mathbf{k}} = \sqrt{q^2+4k^2-4qkx}, 
  \qquad \abs{\mathbf{q}-\mathbf{K}/2} = \sqrt{q^2+K^2/4-qKy}, 
  \qquad
 \abs{\mathbf{K}/2 \pm \mathbf{k}} = \sqrt{K^2/4 + k^2 \pm Kkz}.
\end{align}
A primed summation means a restriction to $L + S + T$ odd and $L+L'$ even.
Integration over $\int d^3q/(2\pi)^3$ yields the full normalization of $n_{\lambda}^{\tau}(q)$ from the first line in Eq.~\eqref{eq:snmd_partial_waves} while the unitary condition from Eq.~\eqref{eq:UdaggerU} expanded in partial waves enforces cancellation of the second and third terms after changing variables appropriately.

The pair momentum distribution in the LDA assuming spherical symmetry is (with explicit $R$ or $R'$ dependence):
%
\begin{align}
    \label{eq:pmd_partial_waves}
    n_{\lambda}^{\tau \tau'}(\qvec,\mathbf{Q}) &= \frac{1}{2} 
    \int d^3R\, 2 \theta(\kF^{\tau}(R)-\abs{\mathbf{Q}/2 + \mathbf{q}}) 
    \int d^3R'\, 2\theta(\kF^{\tau'}(R') -\abs{\mathbf{Q}/2 - \mathbf{q}}) 
    \notag \\
    & \quad \null - \frac{1}{32}(2\pi)^3\delta^3(\qvec) \delta_{\tau,\tau'}
    \int d^3R\, 2 \theta(\kF^{\tau}(R)-\abs{\mathbf{Q}/2}) 
    \notag \\
    & \quad\null + 2\frac{(2\pi)^3}{4\pi} \int d^3R\,\sum_{L,S,T}\!\!{\vphantom{\sum}}'\, \sum_{J}  (2J+1) \frac{2}{\pi}  
    \melr{q (L S) J T}{\delta U}{q (L S) J T}
    |\braket{\tau \tau'}{T\, \tau+\tau'}|^2
    \notag \\
    & \qquad\qquad \null \times
    \theta(\kF^{\tau}(R)-\abs{\mathbf{Q}/2 + \mathbf{q}}) \theta(\kF^{\tau'}(R)-\abs{\mathbf{Q}/2 - \mathbf{q}})  \notag \\
    & \quad \null +  \frac{(2\pi)^3}{4\pi} \int d^3R\, \sum_{L,L',S,T}\!\!\!\!\!{\vphantom{\sum}}'\,\,  \sum_{J} (2J+1) (\frac{2}{\pi})^2 \int_0^{\infty} dk\, k^2  
    \melr{k (L S) J T}{\delta U}{q (L' S) J T} \melr{q (L' S) J T}{\delta U^{\dagger}}{k (L S) J T}   \notag \\  
     & \qquad\qquad \null\times 
    |\braket{\tau \tau'}{T \,\tau+\tau'}|^2
     \,\theta(\kF^{\tau}(R)-\abs{\mathbf{Q}/2 + \mathbf{k}}) \,\theta(\kF^{\tau'}(R)-\abs{\mathbf{Q}/2 - \mathbf{k}}),
\end{align}
\end{widetext}
where for $|\Qvec| \neq 0$, we average over the angle between $\Qvec$ and $\kvec$ to evaluate the $\theta$ functions in the last term.
The first two lines carry the full normalization of the momentum distribution. 
It is easily verified that an integration over $\int d^3q/(2\pi)^3 \int d^3Q/(2\pi)^3$ of these terms in the LDA followed by integrations over $\Rvec$ and $\Rvec'$  yield $Z(Z-1)/2$ if $\tau=\tau' = 1/2$, $N(N-1)/2$ if $\tau=\tau'=-1/2$, and $NZ/2$ if $\tau=\pm 1/2$ while $\tau'=\mp 1/2$.
The unitary condition from Eq.~\eqref{eq:UdaggerU} expanded in partial waves enforces cancellation of the third and fourth terms after integrating over $\qvec$, for any $\Qvec$. 
This is manifest after switching labels for $\kvec$ and $\qvec$ in the last term.

The angle average of the $\theta$ function in the second term of Eq.~\eqref{eq:snmd_partial_waves} is
\begin{align}
    F_1^{\tau\tau'}(q,k) &= \int_{-1}^1 \frac{dx}{2}\,\theta(\kF^{\tau'}-\abs{\mathbf{q}-2\mathbf{k}}) \notag \\
    &=
  \begin{cases} 
   1 & \text{if } q < \kF^{\tau'} \text{and\ } 2k < \kF^{\tau'} - q\\[8pt]
   \frac{{\kF^{\tau'}}^2 - (q-2k)^2}{8kq}  &  \text{if } q < \kF^{\tau'}\ \text{and\ } \\
   & \quad\kF^{\tau'} - q < 2k < \kF^{\tau'} + q \\[8pt]
   \frac{{\kF^{\tau'}}^2 - (q-2k)^2}{8kq} & \text{if }  \kF^{\tau'} < q < \kF^{\tau}\  \text{and\ } \\
   & \quad q - \kF^{\tau'}  < 2k < q + \kF^{\tau'} \\[8pt]
   0 & \text{otherwise}
  \end{cases}
\end{align}
%
%
The angle average of pairs of theta functions that appear several times in Eqs.~\eqref{eq:snmd_partial_waves} and \eqref{eq:pmd_partial_waves} 
is given  by
\begin{align}
    F_2(Q,k) &= \int_{-1}^{1}\frac{dz}{2}\,
    \theta(\kF^{\tau}-\abs{\Qvec/2+\kvec}) 
                    \theta(\kF^{\tau'} - \abs{\Qvec/2-\kvec})
                    \notag \\
  &= 
  \begin{cases} 
  1 & \text{if } k < \kFmin - \frac{Q}{2}\\[8pt]
  \frac{(\kFmin)^2 - (k - Q/2)^2}{2kQ}  &  \text{if } k < \kFmin + \frac{Q}{2} \text{\ and\ } \\
  & \hspace*{-13pt}\kFmin - \frac{Q}{2} < k < \kFmax - \frac{Q}{2} \\[8pt]
  \frac{(\kFavg)^2 - k^2 - Q^2/4}{kQ}  &  \text{if } \kFmax - \frac{Q}{2} < k  \text{\ and\ } \\
  & \quad k < \sqrt{(\kFavg)^2 - \frac{Q^2}{4}} \\[8pt]
  0 & \text{otherwise} 
  \end{cases}
\end{align}
where
\begin{align}
    \kFmin &\equiv \min(\kF^{\tau}, \kF^{\tau'}) ,\\
    \kFmax &\equiv \max(\kF^{\tau}, \kF^{\tau'}) ,\\
    \kFavg &\equiv \sqrt{\frac{1}{2}\Bigl({\kF^{\tau}}^2 +  {\kF^{\tau'}}^2\Bigr)} .
\end{align}
Extensions of the local density approximation applied here are discussed in Ref.~\cite{Tropiano:2021prep}.

\bibliography{tropiano_bib}

\end{document}

%% file: macros.tex
\newcommand{\mystrut}{\rule{0pt}{0.9\normalbaselineskip}}

\newcommand{\singletS}{\ensuremath{^1\mbox{S}_0}}
\newcommand{\tripletS}{\ensuremath{^3\mbox{S}_1}}
\newcommand{\tripletD}{\ensuremath{^3\mbox{D}_1}}

\newcommand{\fmi}{\ensuremath{\mbox{fm}^{-1}}}

\newcommand{\NNLO}{N$^2$LO}

\newcommand{\Uhat}[0]{\widehat U}
\newcommand{\Uhatlam}{\Uhat_{\lambda}^{\phantom{\dagger}}}
\newcommand{\Uhatlamdag}{\Uhat_{\lambda}^{\dagger}}

\newcommand{\delUtilde}[0]{\delta \widetilde U}  

\newcommand{\Ihat}[0]{\widehat I}
\newcommand{\calG}[0]{\widetilde{\mathcal{G}}}

\newcommand{\qvec}[0]{\mathbf{q}}
\newcommand{\kvec}[0]{\mathbf{k}}
\newcommand{\Kvec}[0]{\mathbf{K}}
\newcommand{\Qvec}[0]{\mathbf{Q}}

\newcommand{\rvec}[0]{\mathbf{r}}

\newcommand{\svec}[0]{\mathbf{s}}
\newcommand{\Rvec}[0]{\mathbf{R}}

\newcommand{\khat}[0]{\mathbf{\widehat k}}

\newcommand{\psihat}{\widehat\psi}

\newcommand{\sg}[1]{\sigma_{{#1}}}

\newcommand{\ta}[1]{\tau_{{#1}}}

\newcommand{\half}[0]{\frac{1}{2}}

\newcommand{\akdag}[0]{a^{\dagger}}
\newcommand{\ak}[0]{a^{\phantom{\dagger}}}

\newcommand{\kF}{k_{\scriptscriptstyle{\text F}}}
\newcommand{\kFmin}{\kF^{\text{min}}}
\newcommand{\kFmax}{\kF^{\text{max}}}
\newcommand{\kFavg}{\kF^{\text{avg}}}

\newcommand{\nhat}{\widehat{n}}
\newcommand{\nhathi}{\nhat_{\infty}} 
\newcommand{\nhatlo}{\nhat_{\lambda}}  

\newcommand{\Striplet}{$^3$S$_1$}
\newcommand{\Dtriplet}{$^3$D$_1$}

\newcommand{\rhoSL}{\rho_{\rm SL}}
\newcommand{\rhoDM}{\rho_{\rm DM}}

\newcommand{\melr}[3]{(#1|#2|#3)}